\documentclass[twocolumn,secnumarabic,amssymb, nobibnotes, aps, prc, superscriptaddress,nobalancelastpage,nofootinbib]{revtex4-1}
\usepackage{bm} 
\usepackage{amsmath} 
\usepackage{braket} 
\usepackage{mathtools}
\usepackage[normalem]{ulem}
\usepackage{multirow}
\usepackage{enumitem}
\usepackage{subfig}
\usepackage{graphicx}
\usepackage{booktabs}
\usepackage{placeins}
\usepackage{hyperref}
\usepackage[dvipsnames]{xcolor}
\hypersetup{breaklinks=true,colorlinks=true,linkcolor=blue,citecolor=blue,filecolor=magenta,urlcolor=cyan}

\newcommand{\ca}{$^{48}$Ca}
\newcommand{\Pb}{$^{208}$Pb}
\newcommand{\apv}{A_{PV}}
\newcommand{\pol}{\alpha_{D}}
\newcommand{\nsat}{n_{sat}}
\newcommand{\esat}{E_{sat}}
\newcommand{\ksat}{K_{sat}}
\newcommand{\qsat}{Q_{sat}}
\newcommand{\zsat}{Z_{sat}}
\newcommand{\lsym}{L_{sym}}
\newcommand{\esym}{E_{sym}}
\newcommand{\ksym}{K_{sym}}
\newcommand{\qsym}{Q_{sym}}
\newcommand{\zsym}{Z_{sym}}
\newcommand{\ncc}{n_{cc}}
\newcommand{\lk}{\mathcal{L}}

\begin{document}

\title{Emulator-Assisted Nuclear DFT Inference and Its Consequences\\for the Structure of Neutron Stars}

\author{Pietro Klausner}
\email{pietro.klausner@unimi.it}
\affiliation{Universit\'e de Caen Normandie, CNRS/in2p3, LPC Caen (UMR6534), 14050 Caen, France}
\affiliation{Dipartimento di Fisica ``Aldo Pontremoli'', Universit\`a degli Studi di Milano, 20133 Milano, Italy}
\affiliation{INFN, Sezione di Milano, 20133 Milano, Italy}

\author{Marco Antonelli}
\email{antonelli@lpccaen.in2p3.fr}
\affiliation{Universit\'e de Caen Normandie, CNRS/in2p3, LPC Caen (UMR6534), 14050 Caen, France}

\author{Gianluca Col\`o}
\email{gianluca.colo@mi.infn.it}
\affiliation{Dipartimento di Fisica ``Aldo Pontremoli'', Universit\`a degli Studi di Milano, 20133 Milano, Italy}
\affiliation{INFN, Sezione di Milano, 20133 Milano, Italy}

\author{Francesca Gulminelli}
\email{gulminelli@lpccaen.in2p3.fr}
\affiliation{Universit\'e de Caen Normandie, CNRS/in2p3, LPC Caen (UMR6534), 14050 Caen, France}
\affiliation{Institut Universitaire de France, Paris, France}

\author{Xavier Roca-Maza}
\email{xavier.roca.maza@fqa.ub.es}
\affiliation{Dipartimento di Fisica ``Aldo Pontremoli'', Universit\`a degli Studi di Milano, 20133 Milano, Italy}
\affiliation{INFN, Sezione di Milano, 20133 Milano, Italy}
\affiliation{Departamento de F\'isica Qu\`antica i Astrof\'isica, Mart\'i i Franqu\'es, 1, 08028 Barcelona, Spain}
\affiliation{Institut de Ci\`encies del Cosmos, Universitat de Barcelona, Mart\'i i Franqu\'es, 1, 08028 Barcelona, Spain}

\author{Enrico Vigezzi}
\email{enrico.vigezzi@mi.infn.it}
\affiliation{INFN, Sezione di Milano, 20133 Milano, Italy}


\begin{abstract}
Nuclear density functional theory provides a unified description of finite nuclei and bulk nuclear matter, and is widely used to model the neutron star equation of state. However, extrapolations to supra-saturation densities require a quantified treatment of uncertainties arising from parameter estimation and functional choices.
We present an updated Bayesian inference of a Skyrme energy density functional augmented by a flexible meta-model density dependence at high density. 
Nuclear observables are computed using a Gaussian emulator of the publicly available Milano HFBCS-QRPA code, enabling efficient exploration of a high-dimensional parameter space. Relative to previous analyses, we extend the calibration set with isospin-sensitive data, including masses and charge radii along selected Ca and Sn isotopic chains, and updated constraints from giant monopole resonances. 
The resulting posteriors are further constrained by \emph{ab initio} neutron-matter calculations and astrophysical observations, including recent NICER measurements, yielding consistent crust and core properties of catalyzed NS compatible with current constraints. 
Bulk nuclear-matter parameters are well approximated by a multivariate Gaussian with covariance matrix provided for direct reuse, while several finite-nucleus parameters exhibit pronounced non-Gaussianity.
\end{abstract}

\maketitle

\section{Introduction}
 
Nuclear density functional theory (DFT) is among the most suitable frameworks to constrain effective nuclear interactions using laboratory data \citep{Schunck_book}. Its applicability across the nuclear chart, as well as to transport models for heavy-ion collisions and to bulk nuclear matter, makes DFT a natural tool to extrapolate phenomenological information to unknown regions of the isotope table and to density, temperature, and isospin conditions inaccessible to experiments \citep{Grams2025, Grams2026}.
For this reason, most studies of the neutron star (NS) equation of state (EoS) are carried out within the DFT framework, using relativistic (RMF) or non-relativistic (Skyrme, Gogny) energy density functionals (EDFs) whose form and parameters are calibrated to nuclear data \citep{Scurto2024, Char2025, DinhThi2021, Beznogov2024}; see \citep{Roca2018, Kumar2024} for a recent review.
However, EDFs have limitations related to the assumed functional form governing their density dependence; in addition, they have been extensively tested in cases where experimental data allow their calibration \citep{Yuskel2019,Zhang2023,Xu2022}, and the validity of current EDFs for extremely low-density or high-density neutron matter may be questioned \citep{Erler2012}. 
Even in not so extreme cases, a quantified treatment of parameter-estimation uncertainties is needed for astrophysical extrapolation to regions not covered by phenomenological constraints.
 
To address these issues, recent works \citep{Klausner2025,Klausner2025_2} adopted a Bayesian approach to NS EoS parameter estimation. Bayesian inference yields full posterior distributions for the nucleonic EoS model parameters, whereas traditional $\chi^2$ minimization used to construct many popular functionals provides only point estimates, typically under an implicit Gaussian approximation for uncertainties.

In~\citep{Klausner2025,Klausner2025_2}, a broad set of nuclear-structure data, described with relatively simple many-body approaches such as HF-BCS (\textit{HartreeFock--Bardeen-Cooper–Schrieffer}) and RPA (\textit{Random Phase Approximation}), was used for the inference. 
A full exploration of the high-dimensional parameter space was made possible by Gaussian emulators \citep{Rasmussen_book, madai} of the nuclear-structure codes. 
The density dependence of the EDF was taken from the meta-model framework \citep{Margueron2018}, complemented by the gradient and spin-orbit terms of the standard Skyrme functional.

Although this meta-model extension is not as flexible as other approaches, like Gaussian-Processes EoS \citep{Landry2019, Essick2020}, or piecewise polytropes \citep{Read2009, Suleiman2022}, it assures a seamless extension of the EDF to high densities for astrophysical applications.
In fact, it reduces spurious correlations between low- and high-density regimes \citep{Xu2022} induced by the restrictive Skyrme density dependence. At the same time, a mapping between the meta-model and Skyrme bulk terms at low density preserves the physical correlations between bulk and surface parameters relevant for the NS crust, enabling a consistent determination of the crustal EoS informed by nuclear-structure data.

In this paper, we present a revised and extended version of the inferences in~\citep{Klausner2025,Klausner2025_2}. The main advance is a new inference that exploits more complete information of the masses and charge radii along selected isotopic chains, providing more reliable constraints on the symmetry energy. We update the experimental input of the IsoScalar Giant Monopole Resonance (ISGMR) excitation energy based on the most recent analyses of giant monopole resonances.
Finally, we propagate the resulting posterior distributions to predictions of static NS properties and incorporate the latest NICER constraints, with the goal of further restricting the high-density behavior of the EoS.
 
The paper is organised as follows. Section~\ref{sec:setup} describes the setup of our emulator-assisted inference, with emphasis on the improvements relative to the previous study in~\citep{Klausner2025}. 
Updated results for nuclear matter and Skyrme parameters are presented in Sec.~\ref{sec:nuclear_inference_results}, where we highlight the impact of the isotopic chain information on the symmetry energy parameters. Representative predictions for static NS observables are given in Sec.~\ref{sec:astro}. Finally, Sec.~\ref{sec:Gaussian} introduces a multivariate Gaussian approximation of our full posterior; the corresponding parameters (mean vector and covariance matrix) are reported in the Appendix for use in future analyses.

\section{Inference set up}
\label{sec:setup}

We perform a Bayesian inference using the standard Skyrme ansatz \citep{Chabanat1997} as base model and a rich pool of observables, ranging from ground state and nuclear response, as constraints.
Our method follows the methodology in \citep{Klausner2025}, and we refer to Sec.~II therein for the details.

In Tab. \ref{tab:prior} we show the prior distribution of the model parameters.
The first ten paramaters in the Table are in a one-to-one correspondence with the standard Skyrme ones; see for details \citep{Chen2009,Chen2010}.
The only difference with respect to \citep{Klausner2025} is that  open shell nuclei (OS) will be now part of the observable pool.
To describe the pairing correlations that emerge in OS nuclei, a standard local pairing interaction $v(\vec{r},\vec{r}')=v_0\delta(\vec{r}-\vec{r}')$ is added in the particle-particle channel \citep{RingSchuck}, and  an extra parameter is added to the inference, the pairing strength $v_0$, considering up to 6 levels above the Fermi energy.

\begin{table}
\centering
\caption{Lower and upper limits of the interval of the uniform prior distribution of each model parameter.
}
\label{tab:prior}
\begin{tabular}{lcrr}
\hline
\qquad\qquad\qquad & \quad \multirow{2}{3em}{Units} \quad& \quad Lower\quad  & \quad Upper\quad  \\
&& limit & limit \\
      \hline
      $n_{sat}$ &[fm$^{-3}$]   &   0.150 &    0.175 \\ 
      $E_{sat}$ &[MeV]       & -16.50 &  -15.50 \\ 
      $K_{sat}$ & [MeV]      & 180.00 &  260.00 \\ 
      $E_{sym}$ & [MeV]   &  24.00 &    40.00 \\ 
      $L_{sym}$ &[MeV]         & -20.00 &  120.00\\ 
      $G_0$ &[MeV fm$^5$]&  90.00 &  170.00\\ 
      $G_1$ &[MeV fm$^5$]& -90.00 &   70.00\\ 
      $W_0$ &[MeV fm$^5$] &  60.00 &  190.00\\ 
      $m_0^*/m$&  -  &   0.70 &    1.10\\ 
      $m_1^*/m$&  -  &   0.60 &    0.90\\ 
      $ v_0$ & [MeV fm$^3$] & 150 & 350 \\
\hline
\end{tabular}
\end{table}

The total set of data used in the inference is presented in Tab. \ref{tab:obs}. With respect to \citep{Klausner2025}, all the observables linked to $N=Z$ nuclei have been removed. 
These nuclei, which are generally quite hard for EDFs to reproduce, are very sensitive to poorly known channels of the interaction such as proton-neutron pairing, that are often mocked up by the so-called Wigner term \citep{Satula1997,Satula1997_2,Satula2000}. In the absence of such terms, the inclusion of $N=Z$ nuclei could bias the estimation of the other parameters.
Therefore, to avoid adding extra parameters to our model, we simply removed these observables.
We then added several ground state properties, namely binding energies $B.E.$s \citep{Wang_2021} and charge radii $R_{ch}$ \citep{ANGELI2013} of open shell isotopes of Ca and Sn. To adjust the pairing strength, we also included the neutron pairing gap of $^{120}$Sn, obtained by the odd-even-staggering three-points mass difference formula \citep{Satula1998}, centered in the odd-systems.
Finally, we updated the excitation energy of the giant monopole resonance of $^{90}$Zr: a newer analysis of the raw data found a higher value for the excitation energy, $E_{GMR}^{IS} = 18.65\pm 0.17$ MeV \citep{Gupta2018}, to be compared to the previous value $E_{GMR}^{IS} = 17.66\pm 0.07$ MeV~\citep{Gupta2016}.
As we demonstrate in the next section, the primary effect of a larger excitation energy $E_{GMR}^{IS}$ is to shift the distribution of $\ksat$ and $\nsat$; cf. Fig.~5 of~\citep{Klausner2025}. Concerning the other observables, we refer to \citep{Klausner2025} for their description and associated references.

For all observables, the uncertainties quoted in Tab. \ref{tab:obs} that will enter the likelihood (a product of uncorrelated Gaussian functions) of the Bayesian inference comprise both the experimental uncertainty and the theoretical uncertainty associated to the DFT description. 
The latter is the dominant source of uncertainty for all observables except the isovector properties.
For these observables, we employed directly the experimental error, which is large.
The theoretical errors were chosen to reflect the fact that standard EDFs usually describe binding energies and charge radii within a few MeV and a few $10^{-2}$ fm, respectively, rather than within the much smaller experimental uncertainties. 
While dedicated mass models can describe binding energies across the nuclear chart with a root mean square error of about 0.5 MeV, we adopt broader values, namely 2 MeV for binding energies and 0.05 fm for charge radii, which are typical uncertainty scales for standard EDFs with a limited number of parameters. 
These choices should therefore be understood as effective global model-discrepancy scales (i.e., they set the level of accuracy that we demand from the EDF in the present global calibration) and do not provide an explicit treatment of correlations among theoretical residuals, which would require a dedicated covariance discrepancy model (see e.g., \citep{Melendez2019}). 

\begin{table}  
    \caption{Observables used for the inferences.
    The new ones are in the second part of the Table, while the updated data for $^{90}$Zr $E_{GMR}^{IS}$ is in bold font.}
    \label{tab:obs}
    \centering
    \begin{tabular}{cccc}
    \hline
    \multicolumn{4}{c}{Ground-state properties} \\
    \hline
        & $B.E.$ [MeV]   & $R_{ ch}$ [fm]   & $\Delta E_{ SO}$ [MeV] \\ 
    \hline
    $^{208}$Pb& 1636.4 $\pm$    2.0     &    5.50 $\pm$    0.05  & 2.02 $\pm$    0.50  \\ 
    $^{48}$Ca &  416.0 $\pm$    2.0     &    3.48 $\pm$    0.05  & 1.72 $\pm$    0.50  \\ 
    $^{68}$Ni &  590.4 $\pm$    2.0      & - & -\\ 
    $^{132}$Sn& 1102.8 $\pm$    2.0     &    4.71 $\pm$    0.05  & - \\ 
    $^{90}$Zr & 783.9 $\pm$     2.0     &    4.27 $\pm$    0.05  & - \\
    \end{tabular}\\
    \vspace{0.3cm}
    \begin{tabular}{ccc}
    \hline
    \multicolumn{3}{c}{Isoscalar resonances} \\
    \hline
        & $E_{ GMR}^{ IS}$ [MeV]  & $E_{ GQR}^{ IS}$ [MeV] \\
    \hline
    $^{208}$Pb&   13.5 $\pm$    0.5            &  10.9 $\pm$    0.5 \\
    $^{90}$Zr &   $\mathbf{18.7 \pm    0.5}$   & - \\
    \end{tabular}\\
    \vspace{0.3cm}
    \begin{tabular}{cccc}
    \hline
    \multicolumn{4}{c}{Isovector properties} \\
    \hline
        &   $ \alpha_{ D}$ [fm$^{3}$]  &   $m$(1) [MeV fm$^2$]  & $A_{ PV}$ (ppb) \\
    \hline
    $^{208}$Pb&   19.60 $\pm$    0.60    & 961 $\pm$   22    &     $550   \pm       18$\\
    $^{48}$Ca &    2.07 $\pm$    0.22    & -                 &     $2668  \pm      113$\\
    \hline
    \end{tabular}\\
    \vspace{0.3cm}
    \begin{tabular}{cccc}
    \hline
    \multicolumn{4}{c}{New Data from OS nuclei} \\
    \hline
        & $B.E.$ [MeV]   & $R_{ch}$ [fm]   &   $ \Delta_n$ [MeV]   \\ 
    \hline
    $^{50}$Ca  &  427.5 $\pm$  2.0  & 3.52 $\pm$ 0.05  & - \\ 
    $^{46}$Ca  &  398.8 $\pm$  2.0  & -                & - \\
    $^{44}$Ca  &  381.0 $\pm$  2.0  & -                & - \\ 
    $^{42}$Ca  &  361.9 $\pm$  2.0  & -                & - \\ 
    $^{120}$Sn & 1020.5 $\pm$  2.0  & 4.65 $\pm$ 0.05  & 1.3 $\pm0.2$ \\ 
    $^{112}$Sn &  953.5 $\pm$  2.0  & -                & - \\ 
    $^{124}$Sn & 1050.0 $\pm$  2.0  & -                & - \\ 
    \hline
    \end{tabular}
 \end{table}

To allow the full exploration of the parameter space within acceptable computation time limits, we use the Gaussian emulator \citep{Rasmussen_book}, developed by the MADAI collaboration \citep{madai}, to emulate the behavior of the \texttt{hfbcs-qrpa} code \citep{Colo2013,Colo2021}.
For technical details about the Bayesian inference and the validation of the emulator we refer to~\citep{Klausner2025}.

\begin{figure*}
    \centering
    \includegraphics[width=1\linewidth]{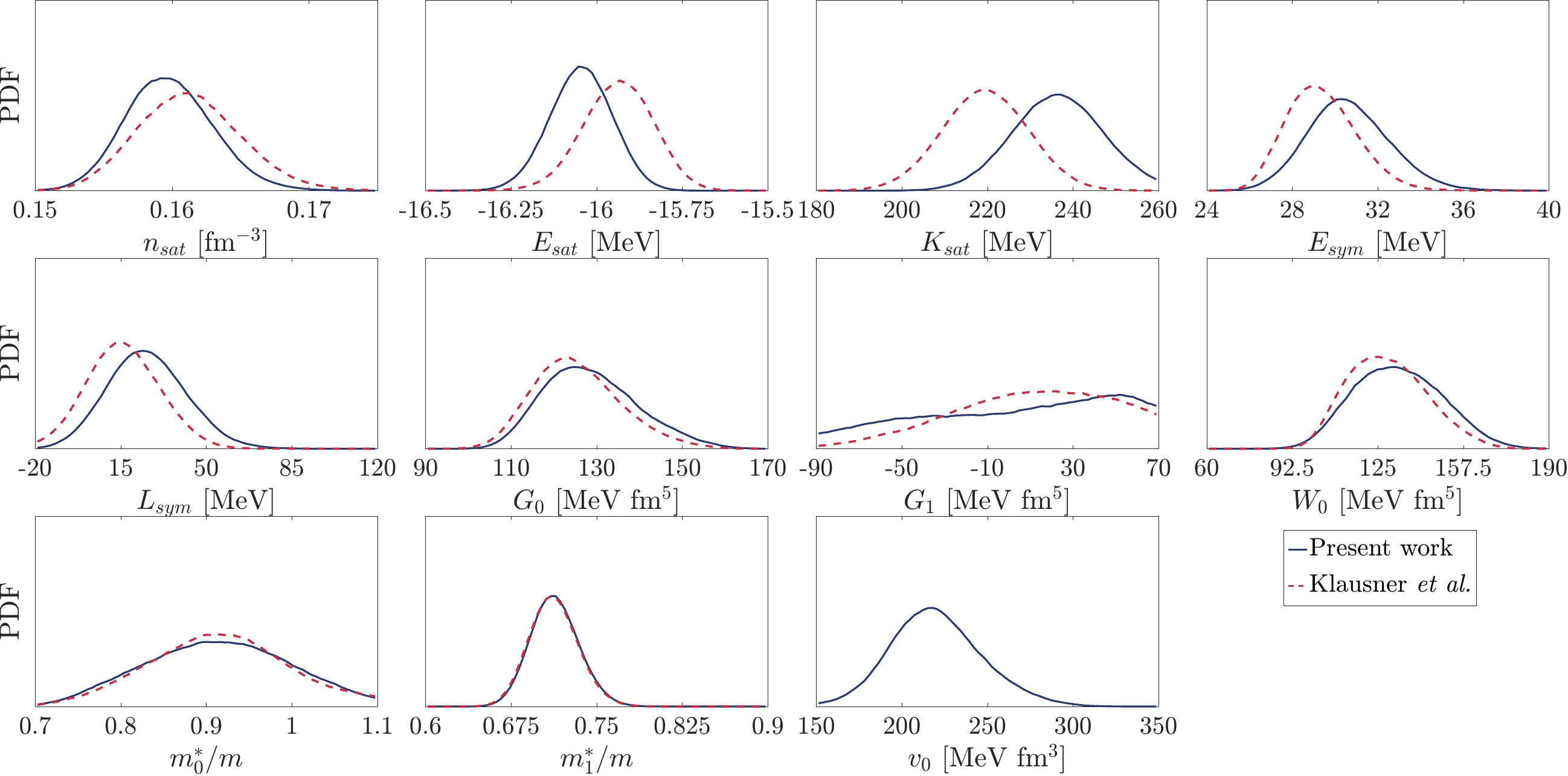}
    \caption{Marginalized posterior distribution of the nuclear matter, surface, spin-orbit, and effective mass parameters  (full blue line), compared to the previous results in~\citep{Klausner2025} (dashed green line).}
    \label{fig:posterior_nuclear_advanced}
\end{figure*}

\section{Inference Results: nuclear structure data}
\label{sec:nuclear_inference_results}

In Fig.~\ref{fig:posterior_nuclear_advanced}, we show the marginalized posterior distribution of the parameters, comparing the new results (full line) with those of \citep{Klausner2025} (dashed line).
The surface parameter $G_0$ and $G_1$, the spin orbit parameter $W_0$ and the two effective masses are only slightly affected by the new data.
$K_{sat}$ is increased as an effect of the new $^{90}$Zr monopole excitation energy and $\nsat$ is lowered as it is anti-correlated to $\ksat$.
Instead, the new calcium and tin isotopes, together with the removal of $N=Z$ nuclei, shift the $\esat$ distribution to lower values.
In general, as we observed in \citep{Klausner2025}, to which we refer for more details, our findings for these parameters are in line with previous studies \citep{Kortelainen2010, Moller2012, Shlomo2006, Khan2012}.
Although both $\esym$ and $\lsym$ are shifted to higher values, however by a rather modest amount, their values are relatively low with respect to other works in the literature~\citep{Roca2018,universe7060182,Kumar2024}. 
Though we already pointed out this feature in our previous analysis \citep{Klausner2025}, we now observe that values of $\lsym > 50$ MeV are less disfavored by the inclusion of open shell nuclei than before, and are not outright excluded anymore. 
This makes our findings more in line with those of previous studies (see, e.g., \citep{Yuskel2019, Zhang2023}).

\begin{figure*}
\centering
    \subfloat[Model results as a function of $L_{sym}$]{
    \includegraphics[width=.95\linewidth]{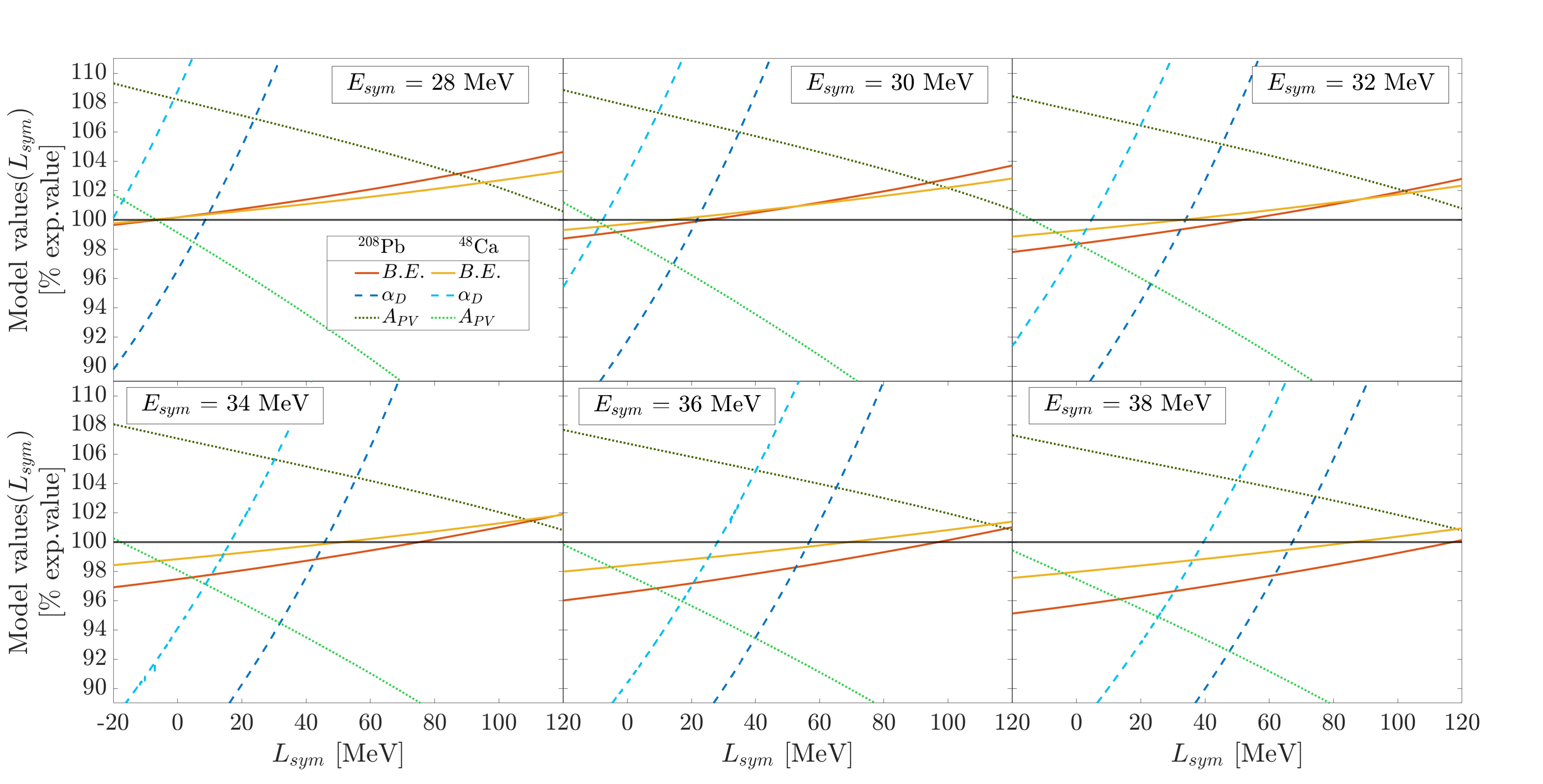}
    \label{fig:L_vs_obs_training_grid}}
    \\
    \subfloat[Likelihood $\lk$ as a function of $L_{sym}$.]{
    \includegraphics[width=.95\linewidth]{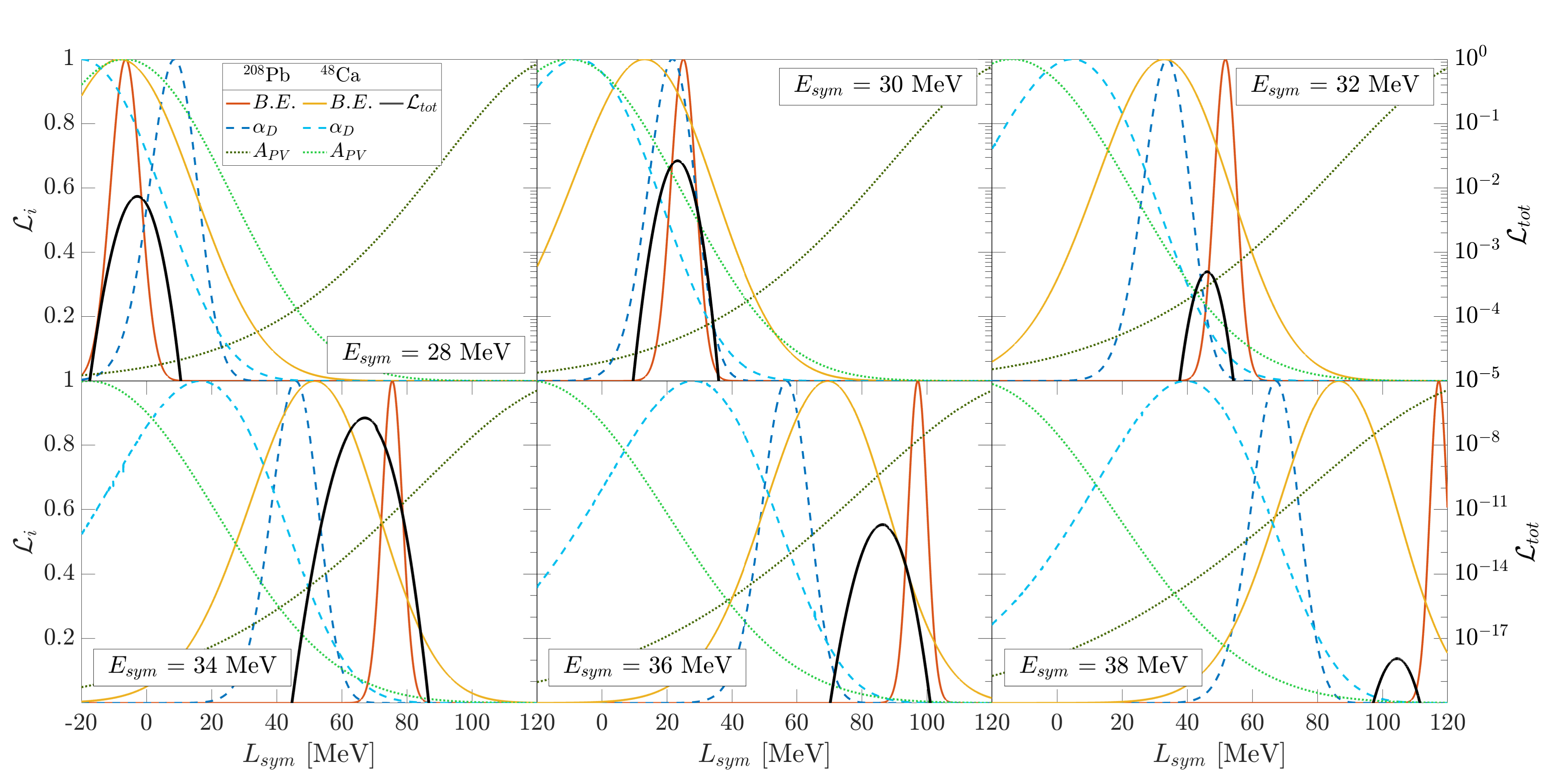}
    \label{fig:L_vs_lik_training_grid}}
    \caption{Model results, Likelihoods $\lk_i$ associated with selected observables in \ca{} and \Pb{} and total Likelihood $\lk_{tot}$ as a function of 
    $L_{sym}$.
    $\esym$ it has been fixed to 6 values, from 28 to 38 MeV in steps of 2 MeV, while all the other parameters have been fixed at the best likelihood model of the complete inference.
    The total likelihood $\lk_{tot}$, product of the single likelihoods $\lk_i$, is shown in logarithmic scale; the corresponding scale is given on the right-hand $y$-axis.}
    \label{fig:L_vs_training_greed}
 \end{figure*}

To investigate this persisting softness of the symmetry energy, we performed a sensitivity study, probing the dependence of some selected observables in \ca{} and \Pb{} on $\lsym$, fixing all the parameters to their best $\log(\lk)$ value, except for $\esym$, which we varied from 28 MeV to 38 MeV, in steps of 2 MeV, and of course $\lsym$, which remained free. 
In Figure \ref{fig:L_vs_obs_training_grid} we show the \texttt{hfbcs-qrpa} code results as a function of $L_{sym}$.
On the $x$-axis we have the full prior interval of $L_{sym}$, while on the $y$-axis the theoretical result of the \texttt{hfbcs-qrpa} code for selected binding energies and for the isovector observables of \ca{} and \Pb{}, expressed as a percentage of the experimental value.
100\%, perfect accordance with the experiment, is highlighted with a black horizontal line.
The orange lines are the  $B.E.$s of the representative nuclei \Pb{} (darker) and \ca{} (lighter).
We can observe the expected correlation induced on $E_{sym}-L_{sym}$:
with increasing $\esym$, a higher value of $\lsym$ is needed to match the experimental result.
Overall, the $B.E.$s vary by about 5\% of the experimental value over the $\lsym$ range.

The dashed lines represent the polarizability $\pol$ of \Pb{} (dark blue) and \ca{} (light blue), while the dotted ones are the parity-violating asymmetries $\apv$ of \Pb{} (dark green) and \ca{} (light green).
As expected from previous works \citep{Reinhard2022},  $\apv$  decreases as $L_{sym}$ grows, while the polarizabilities grow with $\lsym$ \citep{Roca-Maza2013_2,Roca-Maza2015}.
 $\apv$ changes rather slowly, spanning $\approx10-15\%$ of the experimental value across the $\lsym$ prior interval, while $\pol$ has a quite steep gradient, varying significantly in an interval of $\approx40$ MeV. 
If we look at the observables uncertainties in Table \ref{tab:obs}, we see that, while $\apv$ and $\pol$ have a relative error of $\approx$3\% ($\approx$10\% \ca{}), the $B.E.$s impose a much tighter constraint, as they have an error of some parts per thousand.
This brings us to the main argument: 
the posterior will be non negligible only in zones where the model prediction for the $B.E.$s is close to the experimental value.
In $E_{sym}-L_{sym}$ terms, it means that it follows the correlation imposed by the masses.
On that correlation, the optimal solutions can either land on high $\esym-\lsym$ zones, where one can describe fairly well the $B.E.$s together with the \Pb{} $\apv$, or on low $\esym-\lsym$ regions, where one can describe the $B.E.$s and $\pol$ simultaneously.
Since $\pol$ has a much steeper dependence on $\lsym$ than $\apv$, and their relative error is similar, the associated likelihood $\lk_{\alpha_{D}}$ will fall off much more rapidly than $\lk_{A_{PV}}$ when moving away from the regions where the model reproduces the experimental results. 

This becomes even clearer if we look directly at the likelihood $\lk_i$ for each observable.
Figure \ref{fig:L_vs_lik_training_grid} is the same as \ref{fig:L_vs_obs_training_grid}, but instead of showing the model results as a function of $L_{sym}$, it shows directly the $\lk_i$ relative to each observable.
The thick black line is the total likelihood $\lk_{tot}$, i.e., the product of the single likelihoods of the selected observables. 
We see that for $E_{sym}=30$ MeV  $\log(\lk_{tot})$ reaches its highest value; 
at $\esym=28$ MeV its peak decreases by approximately 70\%.
For $E_{sym} \geq  32$ MeV, $\lk_{tot}$ plummets: in the bottom row, where we had to change the scale, we can observe it is several orders of magnitude lower than at $\esym=30$ MeV.
 
Therefore, the reason we find relatively low values of $\esym$ and $\lsym$ with respect to other works in the literature \citep{Roca2018,universe7060182,Kumar2024} lies in the request of simultaneously describing the polarizabilities and the binding energies within the corresponding uncertainties.

Given the notable effect $B.E.$s can have on the $\esym-\lsym$ distributions, adding open shell isotopes of tin and calcium lowers the risk of a singular $B.E.$ biasing the final results. 
However, the arguments we presented in Figure \ref{fig:L_vs_training_greed} still hold, thus explaining the low values of $\esym-\lsym$ we again find in this work.
We can see that a careful choice of the theoretical errors is fundamental to obtain a reliable estimation of the parameters. 
Imposing larger uncertainties to the masses would have resulted in different predictions for the NMP of the isovector sector. 
The uncertainties here considered are consistent with the predictive value of the model, as it can be {\it a posteriori} verified looking at the posterior distribution of $B.E.$s.

\begin{figure*}
    \centering
    \includegraphics[width=0.99\linewidth]{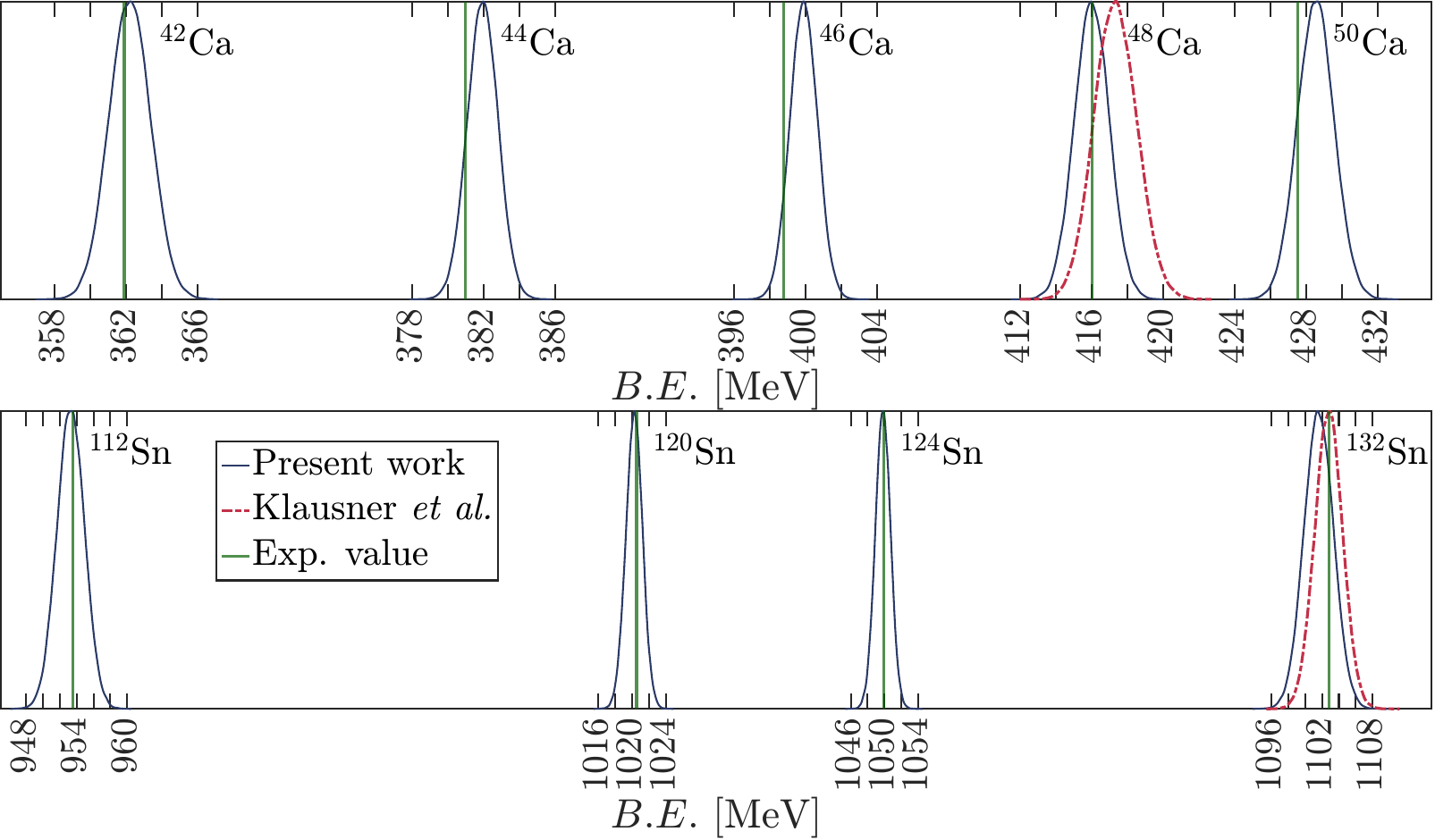}
    \caption{Binding energy posterior distribution for the Ca and  Sn isotopic chains (full blue line), compared to the results of \citet{Klausner2025} (dashed green line).
    The red vertical line marks the experimental value.}
    \label{fig:BE_Ca_Sn}
\end{figure*}

To this aim, we extracted $10^5$ samples from the posterior distribution.
As an example, we display in Fig.~\ref{fig:BE_Ca_Sn} the binding energy posteriors for the Ca and Sn isotopic chains: with the addition of the OS data, we observe a clear improvement of the description of \ca{}, at the price of only a slight worsening of that of $^{132}$Sn.
Given that these distributions are well-behaved and single-peaked, we estimate their width with their standard deviation, which are all in the order of the MeV, as shown in Tab.~\ref{tab:posterior_sigma_BE}.
Therefore, our initial guess of theoretical error (2 MeV) is in line with the predictive power of the model.
\begin{table}[h]
\centering
\begin{tabular}{lcccccccccc}
\hline
& $^{42}$Ca & $^{44}$Ca & $^{46}$Ca & $^{48}$Ca & $^{50}$Ca & $^{112}$Sn & $^{120}$Sn & $^{124}$Sn & $^{132}$Sn \\
\hline
$\sigma$ & 1.16 & 0.86 & 0.77 & 0.97 & 1.03 & 1.63 & 1.00 & 0.91 & 1.74 \\
\hline
\end{tabular}
\caption{Standard deviations $\sigma$, in MeV, of the posterior distribution of $B.E.$ in Figure \ref{fig:BE_Ca_Sn}.}
\label{tab:posterior_sigma_BE}
\end{table}

\begin{figure}
    \centering
    \includegraphics[width=0.99\linewidth]{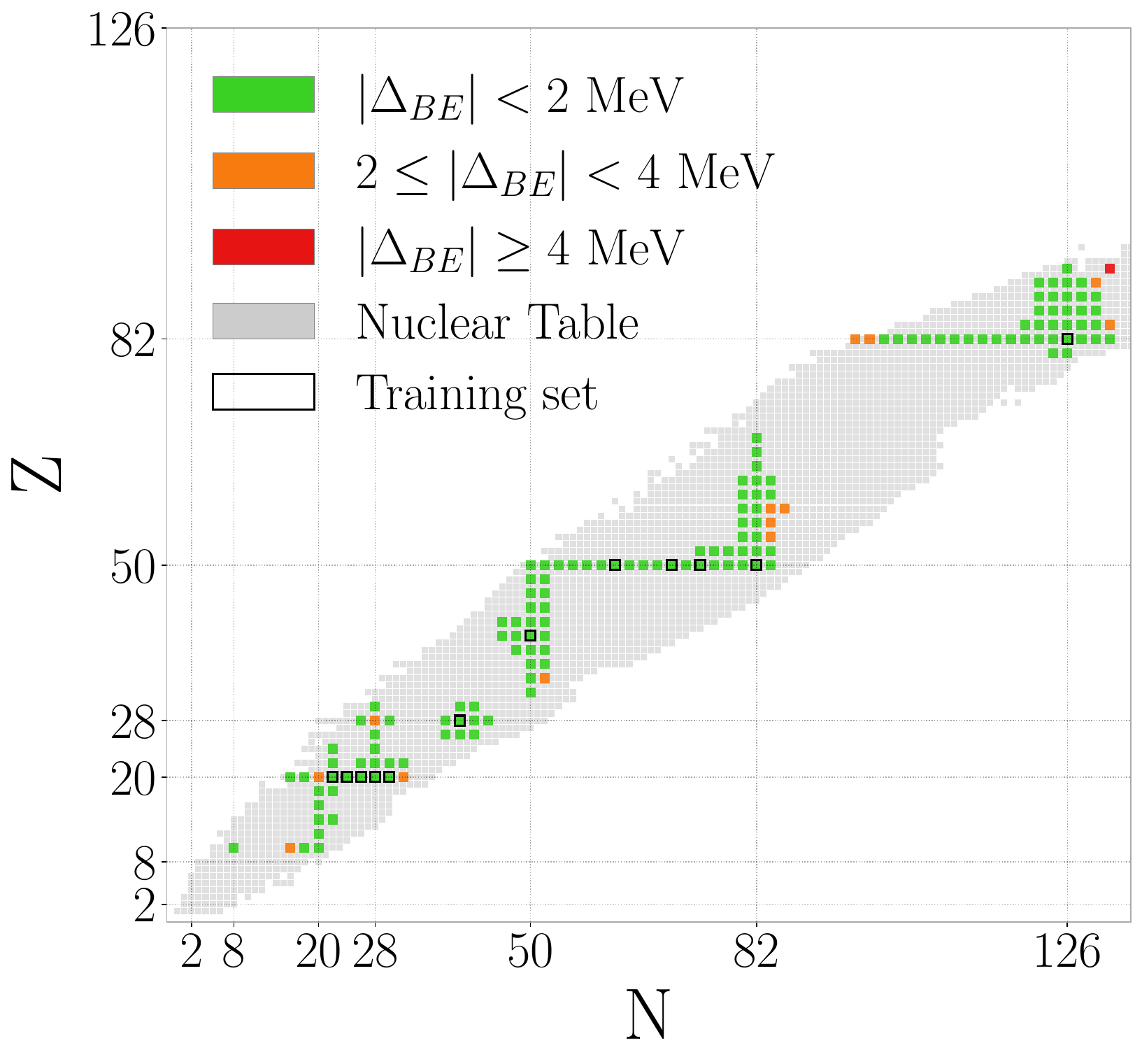}
    \includegraphics[width=0.99\linewidth]{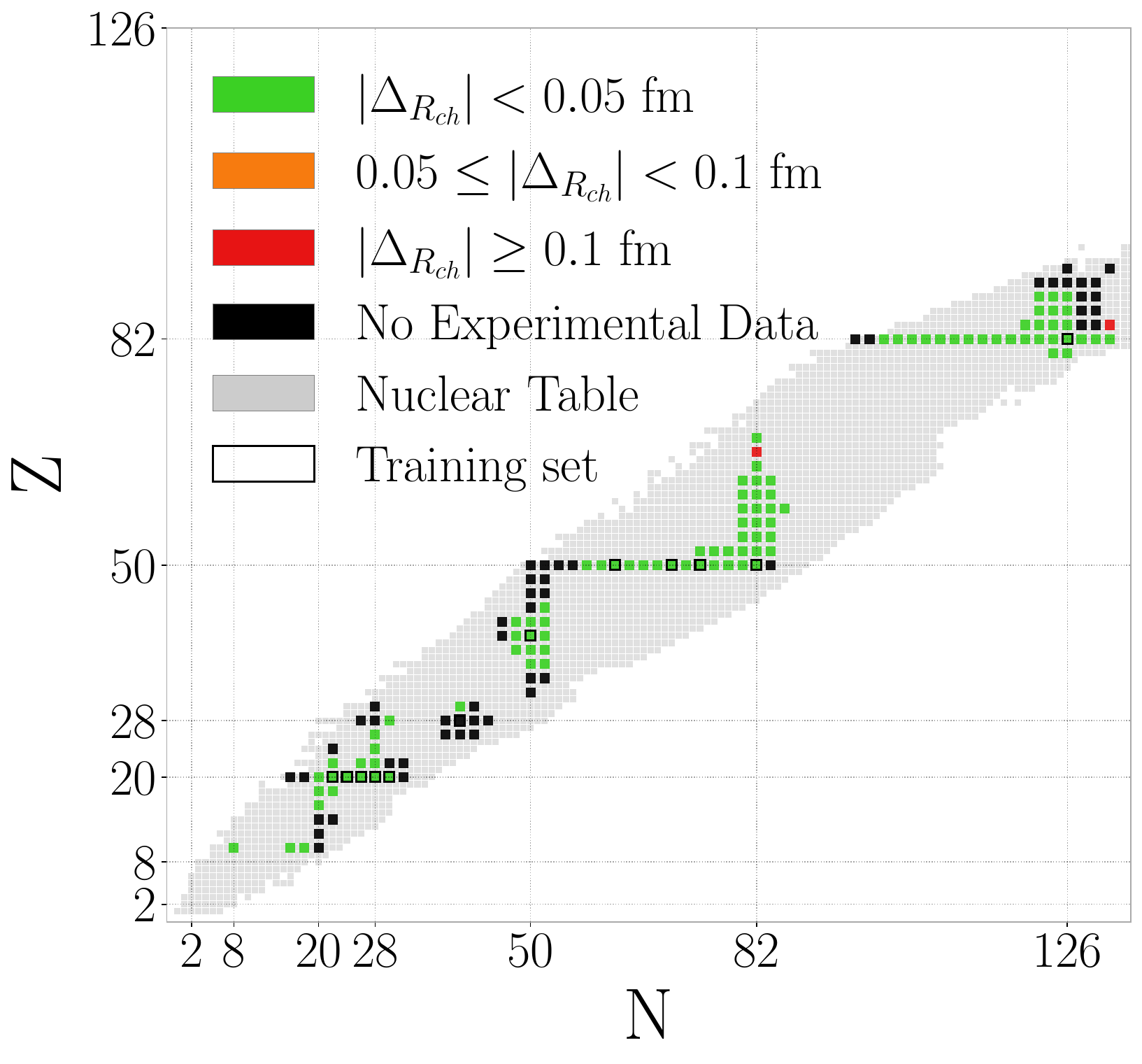}
    \caption{Performance over the nuclear chart (gray squares) of the posterior distribution, for the $B.E.$s of 150 spherical nuclei and $R_{ch}$. 
    The squares relative to different species are colored by the distance $|\Delta_{BE}|$ ($|\Delta_{R_{ch}}|$) between the mean of our posterior and the experimental value:
    green if the deviation is less than 2 MeV or 0.05 fm; orange if it is between 2 and 4 MeV or between 0.05 and 0.1 fm; and finally red if it is more than 4 MeV or 0.1 fm.
    The nuclei used in the training set (Table \ref{tab:obs}) are contoured by a black line.
    }
    \label{fig:Nuclear_chart_comparison}
\end{figure}

We also computed the posterior distribution of binding energies and charge radii of 150 known spherical nuclides along the nuclear chart. We show the results in Figure \ref{fig:Nuclear_chart_comparison}. 
The squares relative to different species are colored by the distance $|\Delta_{BE}|$ ($|\Delta_{R_{ch}}|$) between the mean of our posterior and the experimental value:
green if the distance is less than 2 MeV or 0.05 fm, which are the theoretical errors we assumed (see Tab. \ref{tab:obs}); orange if it is between 2 and 4 MeV or between 0.05 and 0.1 fm; and finally red if it is more than 4 MeV or 0.1 fm.
As we can observe, for the $B.E.$s the vast majority is colored in green (91\%); the rest is mostly orange (8\%), with only 1 instance of red. For the $R_{ch}$, all are green except for two red instances.
As an overall measure of performance, the root-mean-square (RMS) is 1.23 MeV for binding energies and 0.02 fm for charge radii, while the mean distance from experimental values is 0.97 MeV for binding energies and 0.02 fm for charge radii.

\begin{figure*}
    \centering
    \includegraphics[width=1\linewidth]{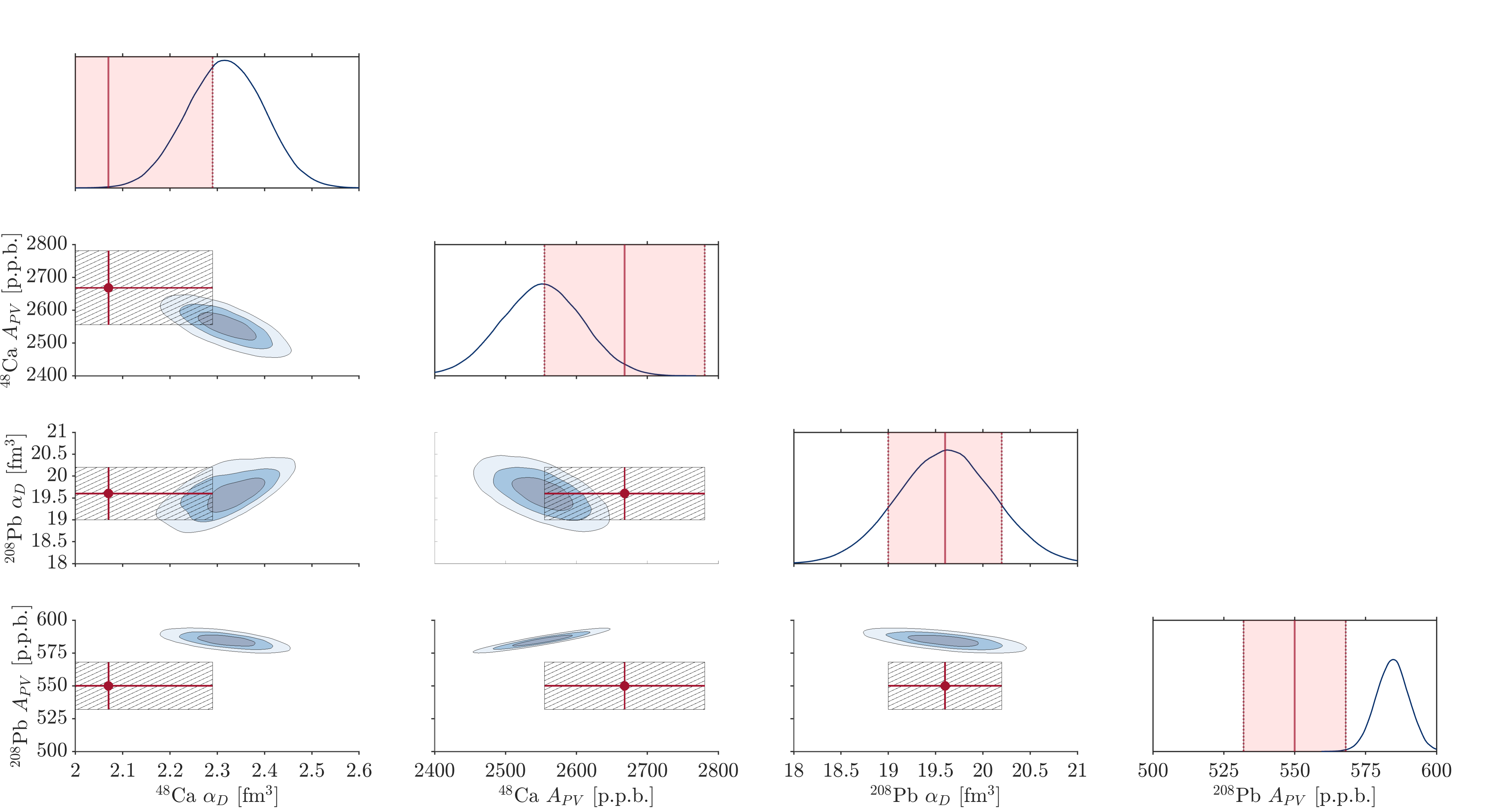}
    \caption{ Corner plot for the $\pol$ and $\apv$ posterior distribution of \Pb{} and \ca{}.
    The red vertical lines in the diagonal plots mark the experimental value, while the red shaded areas show the $1\sigma$ region.
    The rectangular hatched regions of the off-diagonal plots corresponds to the $1\sigma$ uncertainty of both experiments, with the red dot indicating the experimental results.
 }\label{fig:corner_plot_apv_alfaD}
\end{figure*}

Concerning the  isovector-sensitive observables, all  posteriors are compatible with the experimental results, except for the  \Pb{} $\apv$,
in agreement with the results of Ref.~\citep{Klausner2025}. The tension between PREX-II data and the other isovector-sensitive observables is shown in Fig.~\ref{fig:corner_plot_apv_alfaD}, where we plot the corner plot of the $\pol$ and $\apv$ posteriors.
From the marginalized distributions shown in the diagonal plots we can see that \Pb{} $\pol$ is well reproduced by our inference, while \ca{} $\pol$ and \ca{} $\apv$ are slightly over - and under - estimated, but still with significant overlap with the experimental error.
On the other hand, looking at the correlation plots, it is clear that the model cannot reproduce simultaneously \Pb{} $\apv$ and the other isovector-sensitive observables, corroborating other analyses which highlighted this tension. 

The  apparent incompatibility of \Pb{} $\apv$ with the other observables might hint to the need of improving our EDF, that would be important for the description of this specific observable.
For example, it was recently suggested \citep{Yue2024,Zhao2025,Qiu2025} that a strong isovector spin orbit interaction could reconcile the CREX and PREX-II experiments, although following studies hinted that this idea might not be the correct solution to the PREX-CREX puzzle \citep{Athul2025}.
Other strategies, with updated EDF with additional term, were presented in \citep{Salinas2024,Papakonstantinou2026}. 
It would be worth, on the other hand, to measure again the $\apv$ at other facilites and/or with a different kinematics.

The difficulty of reproducing \Pb{} $\apv$ within our theoretical framework naturally leads to question whether the inclusion of this data point in the inference could not lead to biased results.
To answer this question, we have checked that no significative difference is observed in the nuclear matter parameter posteriors when the inference is repeated  omitting the \Pb{} $\apv$ from our pool of data. 
Hence, on the basis of the used EDF,  the incompatibility of the \Pb{} $\apv$ with the other observables in the data pool is not statistically significant.   
 
\section{Inference results: astrophysical data} 
\label{sec:astro}

We now turn to the implications of our results for NS physics. For this analysis, we follow the procedure outlined in~\citep{Klausner2025_2}, to which we refer for details.

We use the full posterior distribution obtained in Sec.~\ref{sec:nuclear_inference_results} as the prior for a subsequent Bayesian inference that includes $\chi$-EFT calculations of pure neutron matter (from 0.02 fm$^{-3}$ to 0.2 fm$^{-3}$) and astrophysical constraints: maximum pulsar masses, tidal polarizability from GW170817, and four independent NICER measurements; see~\citep{monte2025aa,Klausner2025_2} for a precise definition of the associated Bayesian likelihoods. A recent reanalysis suggests a lower mass for the pulsar J0348+0432 (i.e., $1.806 \pm 0.037\,M_\odot$, see \citep{Saffer_2025}) than the previously inferred value of $2.01 \pm 0.04\,M_\odot$ \citep{Antoniadis2013}. Relative to \citep{Klausner2025_2}, we therefore replace the previous maximum observed NS mass constraint from~\citep{Antoniadis2013} with the $2.08 \pm 0.07\,M_\odot$ mass of pulsar J0740+6620 inferred from Shapiro delay measurements~\citep{Fonseca_2021}.
Concerning the other astrophysical and stability-causality constraints, we refer to \citep{monte2025aa,Klausner2025_2} for details.

Using the nuclear posterior from Sec.~\ref{sec:nuclear_inference_results} as the prior for the astrophysical update has two advantages. 
First, it propagates all the correlations among nuclear matter parameters implied by our emulator-assisted calibration. 
Second, it ensures that all parameter sets retained by the combined astrophysics+laboratory inference remain consistent with the nuclear observables used in the first stage fit. 
In particular, because bulk and surface terms are inferred jointly in the first stage fit of Sec.~\ref{sec:nuclear_inference_results}, each parameter set uniquely defines a unified crust-core EoS: the same EDF determines homogeneous matter in the core and all the relevant crust properties, which are computed with the semi-classical extended Thomas-Fermi approach as in~\citep{Klausner2025_2}.
In our hybrid Skyrme + meta-model inference, the nuclear and astrophysical constraints are combined sequentially via factorised likelihoods: finite-nucleus observables constrain the Skyrme sector, and the resulting multidimensional posterior is propagated to the neutron star inference through the Skyrme-derived interface used by the meta-model.
This interface includes the low-order NMPs sampled in the Skyrme sector of the inference (see Table \ref{tab:prior}) and derived quantities such as $K_{sym}^{\rm Sk}$, which is not independently sampled (see \citep{Klausner2025_2} for details). 
The meta-model is then used to provide the additional high-density freedom required for neutron-star matter. 
Therefore the procedure is not an attempt to reproduce a pure Skyrme extrapolation at high density, nor a claim of equivalence with an unavailable fully unified inference, where the same nuclear model is used for both finite nuclei and neutron star matter. 
It is a controlled inference protocol that preserves the nuclear information relevant for the low-density and crust sectors while avoiding the spurious high-density rigidity of the Skyrme functional. 
Moreover, our protocol is expected to retain most of the posterior correlations among the low-order NMPs, as they are directly informed by both astrophysical and nuclear data. 

\begin{figure}
    \centering
    \includegraphics[width=1\linewidth]{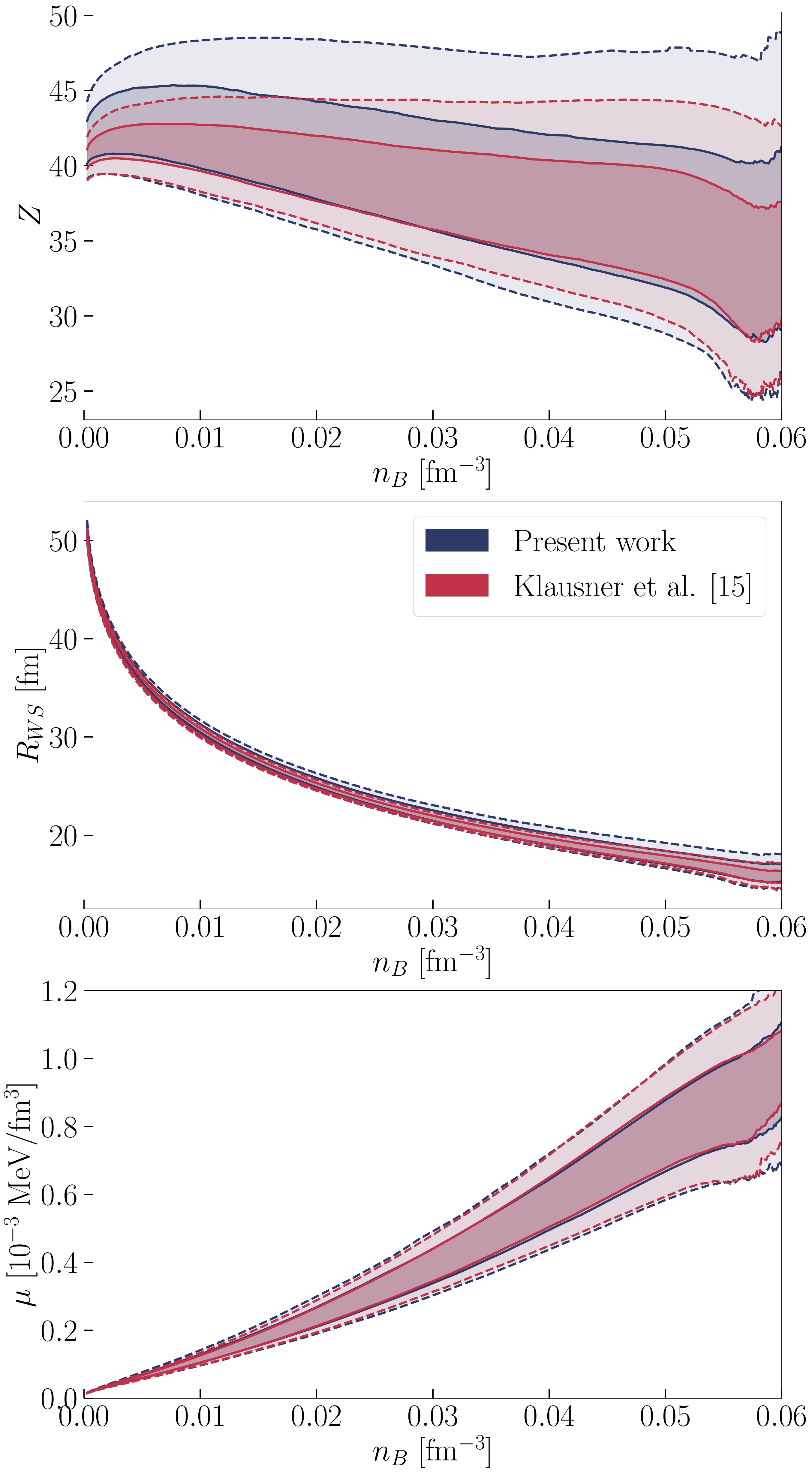}
    \caption{
    Posterior distributions of selected crust properties (proton number in the Wigner-Seitz cell $Z$, cell radius $R_{\rm WS}$, and shear modulus $\mu$) for the two inferences: this work (light blue) and the previous results in~\citep{Klausner2025_2} (light green). Darker shaded regions show the 68\% credible interval, and lighter shaded regions the 95\% credible interval.
    }
    \label{fig:crust_diff_prior}
\end{figure}

The more detailed information on the symmetry sector of the EoS provided by the inclusion of Ca and Sn isotopic chains has a sizeable influence on the crust composition, as can be seen from the upper panel of Fig.~\ref{fig:crust_diff_prior}, which shows the posterior distribution of the ion charge $Z$ as a function of the baryon density in the crust. The higher value of the symmetry energy at low density compared to the results of \citep{Klausner2025} leads to systematically higher ion charges. This is in qualitative agreement with the microscopic calculations of \citep{Pearson2018}, where lower symmetry energies were shown to yield lower proton fractions in the crust, based on HFB calculations with four selected BSK functionals (see also~\citep{RocaMaza2008}). 

On the other hand, the improved estimation of the nuclear matter parameters does not have a sizeable influence on the global properties of the Wigner-Seitz cell, such as the central and gas density (not shown) or the Wigner-Seitz radius, shown in the central panel of Fig.~\ref{fig:crust_diff_prior}. The same is true for the thermodynamic properties of the crust. This includes the crustal pressure (see Fig.~\ref{fig:eos_diff_prior} below) and the shear modulus, displayed in the lower panel of  Fig.~\ref{fig:crust_diff_prior}. This latter is important to estimate dynamic properties of the crust, such as torsional oscillations and interface modes, see e.g.~\citep{Sotani2024}. Taking into account the effect of the ion finite size, the shear modulus can be approximated as~\citep{Zemlyakov2023}:
\begin{equation}
\mu=0.1194  \frac {3(Ze)^2} {4\pi R_{WS}^4}  
\left (1- \frac{u^{5/3}}{2-4u^{1/3}+3u} \right ) \, ,
\end{equation}
$u$ being the cluster volume fraction, see also \citep{Zemlyakov2025}.
From Fig.~\ref{fig:crust_diff_prior} we can see that the modification of the ion charge has a negligible impact on the estimation of the shear modulus.

The influence of the additional information provided by the open shell nuclei is also very weak in the determination of the global stellar properties.
This is demonstrated by Fig.~\ref{fig:eos_diff_prior}, where we show the EoS (pressure $P$ vs mass density $\rho$) and mass-radius posteriors relation of our results in comparison to those of~\citet{Klausner2025_2}.
For the EoS, we also show, in various degree of orange, the EoS of prior samples satisfying the chiral constraint.
Their $\beta-$equilibrated NS EoS is stable (no pressure drop with increasing density) and causal (speed of sound not exceeding that of light).

\begin{figure}
    \centering
    \includegraphics[width=1\linewidth]{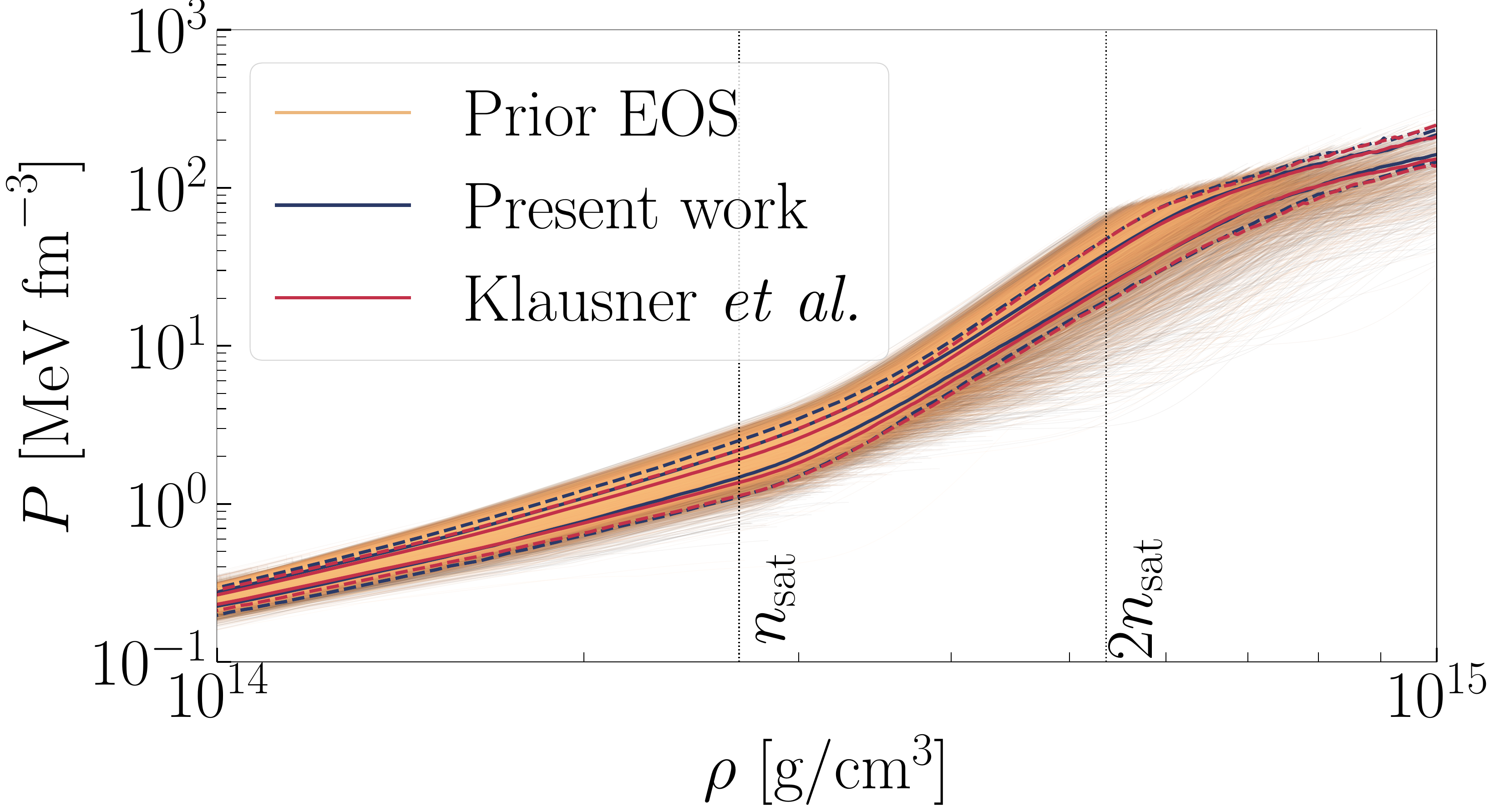}
    \includegraphics[width=1\linewidth]{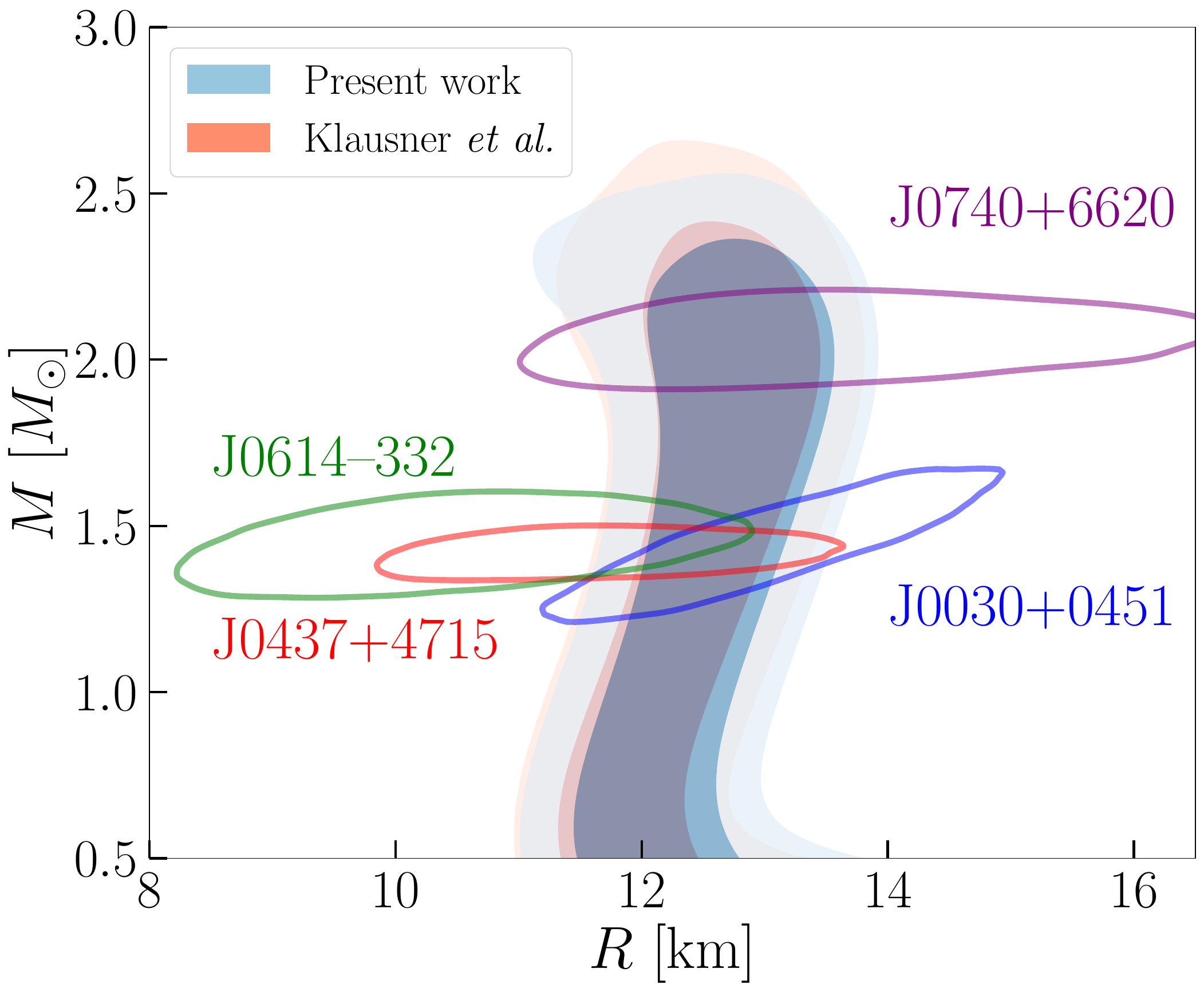}
    \caption{Posterior of the neutron star EoS (upper part) and of the mass-radius relation (lower part) for the two inferences: this work in blue, and our previous results \citep{Klausner2025_2} in red. 
    In the upper panel, the straight line encloses the 68\% CI region, while the dotted the 95\%;
    in the lower, the darkest region encloses the 68\% CI region, while the lighter the 95\%. 
    The four coloured contours in the lower panel encircle the 95\% probability region of the four NICER measurements~\citep{Miller2021,Choudhury2024,Vinciguerra2024,Mauviard2025}.
    For the upper plot, we also plotted in orange prior samples of the neutron star EoS satisfying the chiral constraint.
    } \label{fig:eos_diff_prior}
\end{figure}

As expected, the two results are very similar, and are well compatible with NICER measurements.
The main difference between this work and the results presented in \citep{Klausner2025_2}
concern the nuclear data selected for the analysis, and particularly the revised value of the giant monopole resonance and the 
inclusion of the mass and radius information of $^{42,44,46,48,50}$Ca and $^{112,120,124,132}$Sn isotopes. 
As we have seen in Section \ref{sec:nuclear_inference_results}, this experimental information only affects the low order nuclear matter parameters 
$E_{sat}, K_{sat}, E_{sym}, L_{sym}$. Consistently, the only visible differences of the two analyses concern the low density region of the EoS and the behavior of the mass-radius correlation for masses below $\approx 1 M_\odot$.
The stiffening of the EoS induced by the more complete nuclear physics information of the present work leads to a slight increase of the NS radius for extremely low masses, which however are not expected to be produced in the stellar evolution.

In particular, the slope changes in the EoS between one and two-times saturation densities, already observed in \citep{Klausner2025_2}, is still present in this new version of the analysis. 
Indeed the nuclear structure information suggests a very soft EoS especially in the isovector sector. 
The 2 $M_\odot$ maximum mass requirement therefore demands an important stiffening of the EoS at densities above the ones explored by nuclear structure experiments, and this stiffening is moderated at high density by the gravitational wave GW170817 information.

Finally, we summarize in Tab. \ref{tab:complessive_star_data} other properties of NSs for the two representative masses $M=1.4,\,2\, M_\odot$: their radius $R$, their crustal radius $R_c$, the adimensional tidal deformability $\Lambda$, and the density $\ncc$ and pressure $P_{cc}$ at the crust-core transition.
It is worth noting that we do not find the bimodality feature of the $R_{1.4}$ posterior distribution, as was reported in \citep{Rutherford2024}; rather, our posterior is single--peaked at approximately 12.5 km, and decays to effectively 0 below 11.5 km and above 13.5 km

\begin{table}
\centering
\renewcommand{\arraystretch}{1.5}
\caption{Median value and 68\% confidence interval limits of the star radius $R$, the crust radius $R_c$, the adimensional tidal deformability $\Lambda$, for 1.4 and 2.0 $M_\odot$ NSs, and the density $n$ and pressure $P$ of the crust-core transition.
 All this quantity are coimputed from the nuclear priors described above.}
\label{tab:complessive_star_data}
\begin{tabular}{ccccc}
\hline
&  &$M [M\odot]$    & Present Work       & \citet{Klausner2025}  \\ 
\hline
\multirow{2}{*}{$R$} & \multirow{2}{*}{[km]} & 2.0 & $12.9^{+0.4}_{-0.4}$ & $12.9^{+0.3}_{-0.3}$  \\ 
&  & 1.4 & $12.6^{+0.3}_{-0.3}$ & $12.5^{+0.3}_{-0.3}$  \\ 
\hline
\multirow{2}{*}{$R_c$} & \multirow{2}{*}{[km]} & 2.0 & $0.71^{+0.07}_{-0.07}$ & $0.71^{+0.06}_{-0.06}$  \\ 
&  & 1.4 & $1.16^{+0.09}_{-0.08}$ & $1.14^{+0.07}_{-0.08}$  \\ 
\hline
\multirow{2}{*}{$\Lambda$} & & 2.0 & $68.5^{+18.8}_{-15.2}$ & $67.4^{+16.5}_{-11.9}$  \\ 
&  & 1.4 & $557.6^{+98.5}_{-75.2}$ & $544.8^{+83.1}_{-69.1}$  \\ 
\hline
$n_{cc}$ &[fm$^{-3}$] & - & $0.090^{+0.008}_{-0.007}$ & $0.092^{+0.011}_{-0.009}$  \\ 
$P_{cc}$ & [MeV fm$^{-3}$]  & - & $0.52^{+0.07}_{-0.08}$ & $0.52^{+0.08}_{-0.08}$  \\ 
\hline
\end{tabular}
\end{table}

\section{Gaussian approximation of nuclear posterior}
\label{sec:Gaussian}

The nuclear posterior distribution of Fig.~\ref{fig:posterior_nuclear_advanced}, containing a large amount of available information from nuclear structure data, can be used as a nuclear informed prior for more advanced astrophysical analyses than the simple applications shown in the present paper.
It can also be used  for the generation of representative EoS models complying with nuclear physics constraints, to be used in dedicated astrophysical simulations.

For these future applications, an analytical approximation of the posterior might be useful.
The distributions in Fig.~\ref{fig:posterior_nuclear_advanced} are bell-shaped, except for the parameter $G_1$.
Therefore, a good analytical representation of this posterior is given by a multivariate Gaussian:
\begin{equation}
    \mathcal{N}(\mathbf{x}|\boldsymbol{\mu},\boldsymbol{\Sigma}) \propto
    \exp \left(-\frac{1}{2}(\mathbf{x}-{\boldsymbol{\mu}})^{\mathrm{T}} \boldsymbol{\Sigma}^{-1}(\mathbf{x}-{\boldsymbol{\mu}})\right)
    \label{eq:multivariate_Gaussian}
\end{equation}
where ${\boldsymbol{\mu}}$ is the mean vector and $\boldsymbol{\Sigma}$ is the covariance matrix (both can be easily obtained from our posterior sample and are given in the Appendix).

\begin{figure*}
    \centering
    \includegraphics[width=1\linewidth]{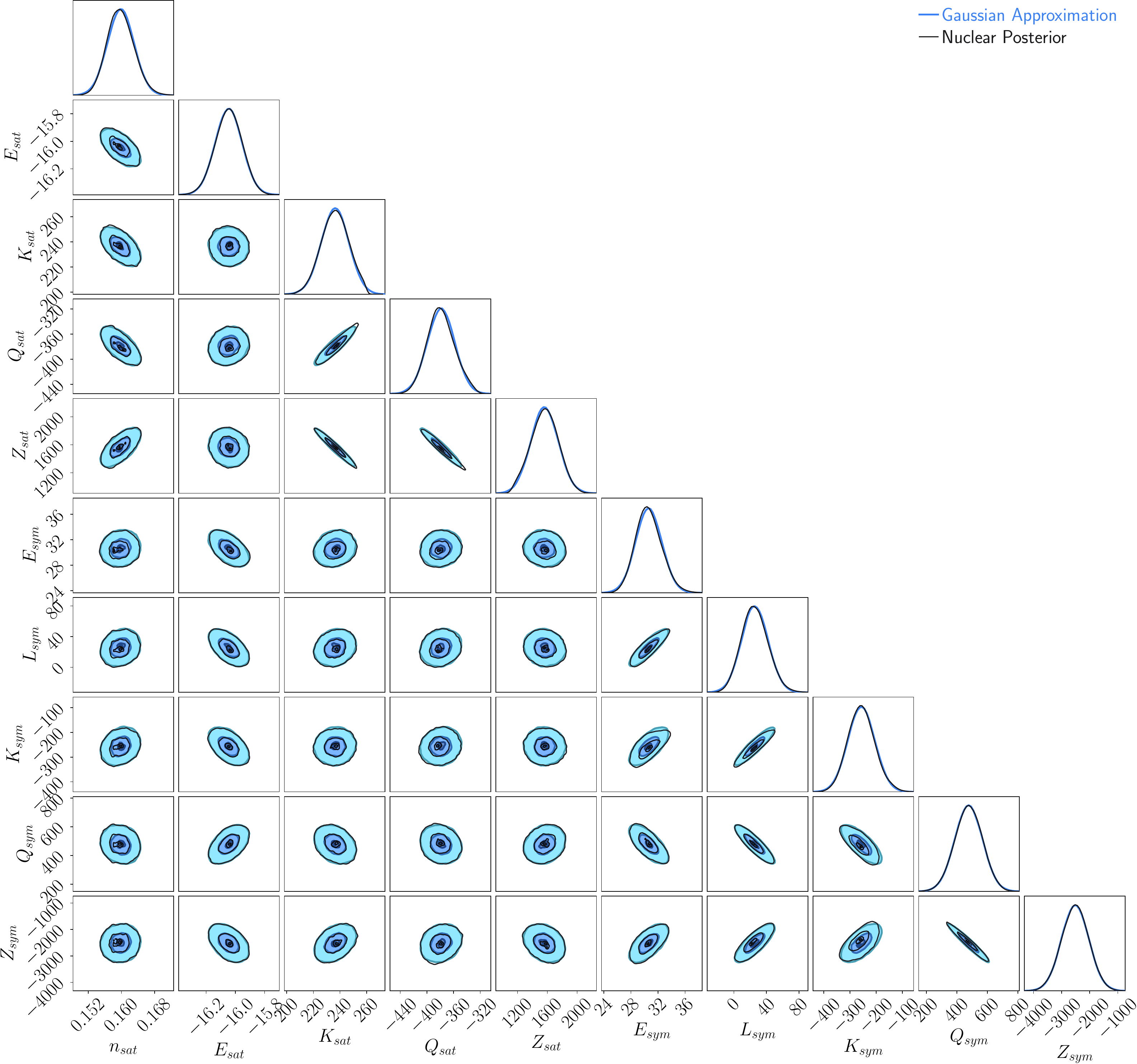}
    \caption{Gaussian approximation of the nuclear posterior. Focus: nuclear matter parameters.}
    \label{fig:prior_vs_gaussian_nmp}
\end{figure*}

\begin{figure}
    \centering
    \includegraphics[width=1\linewidth]{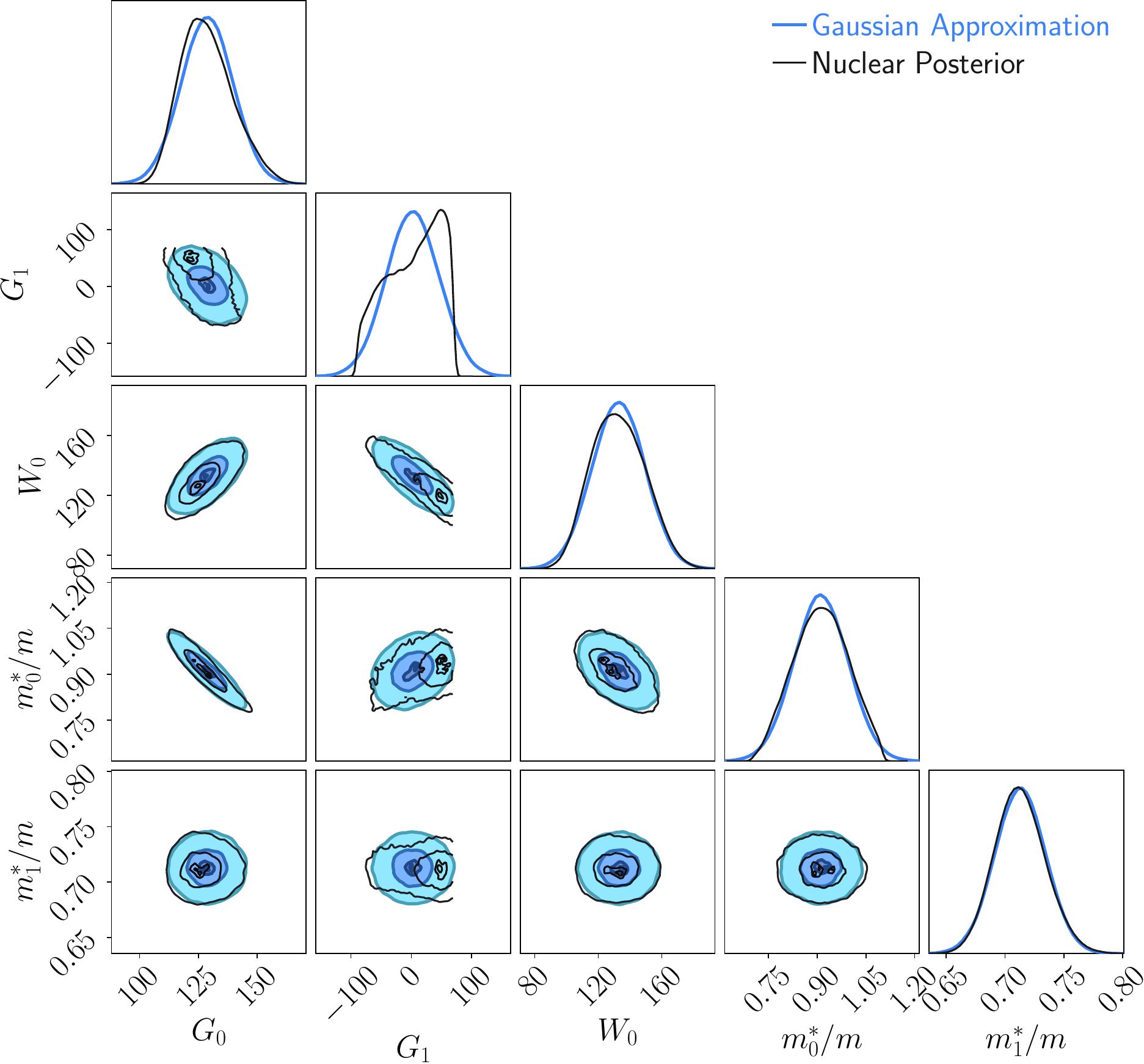}
    \caption{Gaussian approximation of the nuclear posterior. Focus: surface parameters parameters.}
    \label{fig:prior_vs_gaussian_surf}
\end{figure}

Figures~\ref{fig:prior_vs_gaussian_nmp} and~\ref{fig:prior_vs_gaussian_surf} compare the full posterior obtained via our sampling procedure with its analytical approximation $\mathcal{N}(\mathbf{x}|\boldsymbol{\mu},\boldsymbol{\Sigma})$. Apart from $G_1$, whose bimodal behaviour cannot be captured by a single multivalued Gaussian, the remaining parameters are well reproduced.

Regarding Fig.~\ref{fig:prior_vs_gaussian_nmp}, it is important to stress that in standard Skyrme interactions the higher-order nuclear-matter parameters $\qsat$, $\zsat$, $\ksym$, $\qsym$, and $\zsym$ are constrained by the lower-order ones \citep{Xu2022}.
For meaningful astrophysical applications, these parameters should not be treated as independent degrees of freedom above saturation, as discussed in~\citep{Klausner2025_2}.
However, they may be needed for some  applications concerning the sub-saturation EoS, like for studying crust properties.
Therefore, for completeness, we include their distribution in our multivariate Gaussian.

The optimal values of ${\boldsymbol{\mu}}$ and $\boldsymbol{\Sigma}$ are listed in the Appendix, see Tables \ref{tab:multivariate_gaussian_mean} and \ref{tab:multivariate_gaussian_cov}, and are also available in the Supplemental Materials \cite{suppmaterial} in a more convenient format.

\section{Conclusions}

Using a Gaussian emulator of the publicly available Milano \texttt{hfbcs-qrpa} code \citep{Colo2013,Colo2021} for nuclear structure calculations, we have performed a Bayesian analysis of a large set of EoS-sensitive static and dynamic observables, including both closed shell and open shell nuclei. A standard Skyrme interaction augmented with a more flexible density dependence at super-saturation density is employed. 
The posterior 16-dimensional distribution of the associated bulk and surface parameters is used in a second Bayesian inference to include constraints from ab-initio neutron matter calculations and astrophysical observations.
This allows a consistent calculation of crustal as well as core properties of catalized neutron stars, which is in agreement with present observations.
The posterior distribution of bulk parameters 
can be accurately fitted by a multivariate Gaussian, whose parameters are given for future applications. Concerning the parameters characteristic of finite nuclei (surface, spin-orbit and pairing), their complex distribution shows considerable deviations from a Gaussian behavior, and numerical values are available upon request to the authors.

This work is a follow-up of similar previous analyses \citep{Klausner2025,Klausner2025_2}. With respect to these works, we have increased the pool of isospin-sensitive observables by extending the analysis to open-shell nuclei with the inclusion of an extra pairing parameter. 
The extra information brought by the inclusion of Ca and Sn isotopic chains leads to a sizeable modification of the low-order nuclear matter parameters in comparison to \citep{Klausner2025}, and an improved description of the neutron star crust.

Possible extensions of the present work include applying the same inference protocol to different EDF families, such as RMF models, or to more flexible non-relativistic functionals, such as the KIDS \citep{Gil2017, Gil2017_ActaPolonica, Papakonstantinou2018} or Fayans functionals \citep{Fayans1998}, which could provide additional freedom in the isovector sector (see \citep{Xu2022}). 
A systematic comparison of posterior distributions obtained with different functionals, but with the same statistical protocol and the same calibration data, would help quantify the model dependence of the inferred nuclear-matter and neutron-star properties.
A complementary issue concerns the choice of nuclear observables entering the likelihood. 
As discussed above, both the selected dataset and the adopted theoretical uncertainties can have a significant impact on the posterior distributions. 
It would therefore be useful to investigate which minimal set of nuclear observables is most effective in constraining the behavior of nuclear matter, which uncertainties should be assigned to the different classes of observables, and which measurements should be excluded. 
Such studies would help establish more robust protocols for future Bayesian EDF calibrations and would make comparisons among different statistical analyses less ambiguous.

\acknowledgments

This work has been partially supported by the IN2P3 Master Project MAC.
X.R.M. acknowledges support by MI-CIU/AEI/10.13039/501100011033 and by FEDER UE through grants PID2023-147112NB-C22; and through the ``Unit of Excellence Maria de Maeztu 2025–2028'' award to the Institute of Cosmos Sciences, grant CEX2024-001451-M. Additional support is provided by the Generalitat de Catalunya (AGAUR) through grant 2021SGR01095.

\bibliographystyle{apsrev4-1}
\bibliography{bibliography}

@article{universe7060182,
	title        = {Progress in Constraining Nuclear Symmetry Energy Using Neutron Star Observables Since GW170817},
	author       = {Li, Bao-An and Cai, Bao-Jun and Xie, Wen-Jie and Zhang, Nai-Bo},
	year         = 2021,
	journal      = {Universe},
	volume       = 7,
	number       = 6,
	doi          = {10.3390/universe7060182},
	issn         = {2218-1997},
	url          = {https://www.mdpi.com/2218-1997/7/6/182},
	article-number = 182
}

@article{ANGELI2013,
	title        = {Table of experimental nuclear ground state charge radii: An update},
	author       = {I. Angeli and K.P. Marinova},
	year         = 2013,
	journal      = {Atomic Data and Nuclear Data Tables},
	volume       = 99,
	number       = 1,
	pages        = {69--95},
	doi          = {https://doi.org/10.1016/j.adt.2011.12.006},
	issn         = {0092-640X},
	url          = {https://www.sciencedirect.com/science/article/pii/S0092640X12000265}
}

@article{Antoniadis2013,
	title        = {A Massive Pulsar in a Compact Relativistic Binary},
	author       = {{Antoniadis}, John and {Freire}, Paulo C.~C. and {Wex}, Norbert and {Tauris}, Thomas M. and {Lynch}, Ryan S. and {van Kerkwijk}, Marten H. and {Kramer}, Michael and {Bassa}, Cees and {Dhillon}, Vik S. and {Driebe}, Thomas and {Hessels}, Jason W.~T. and {Kaspi}, Victoria M. and {Kondratiev}, Vladislav I. and {Langer}, Norbert and {Marsh}, Thomas R. and {McLaughlin}, Maura A. and {Pennucci}, Timothy T. and {Ransom}, Scott M. and {Stairs}, Ingrid H. and {van Leeuwen}, Joeri and {Verbiest}, Joris P.~W. and {Whelan}, David G.},
	year         = 2013,
	month        = apr,
	journal      = {Science},
	volume       = 340,
	number       = 6131,
	pages        = 448,
	doi          = {10.1126/science.1233232},
	archiveprefix = {arXiv},
	eprint       = {1304.6875},
	primaryclass = {astro-ph.HE},
	adsurl       = {https://ui.adsabs.harvard.edu/abs/2013Sci...340..448A}
}

@article{Athul2025,
	title        = {Role of the isovector spin-orbit potential in mitigating the CREX-PREX dilemma},
	author       = {Kunjipurayil, Athul and Piekarewicz, J. and Salinas, Marc},
	year         = 2025,
	month        = jul,
	journal      = {Phys. Rev. C},
	publisher    = {American Physical Society},
	volume       = 112,
	pages        = {014310},
	doi          = {10.1103/tcy2-brmk},
	url          = {https://link.aps.org/doi/10.1103/tcy2-brmk},
	issue        = 1,
	numpages     = 10
}

@article{Chabanat1997,
	title        = {A Skyrme parametrization from subnuclear to neutron star densities},
	author       = {{Chabanat}, E. and {Bonche}, P. and {Haensel}, P. and {Meyer}, J. and {Schaeffer}, R.},
	year         = 1997,
	month        = feb,
	journal      = {Nuc. Phys. A},
	volume       = 627,
	pages        = {710--746},
	doi          = {10.1016/S0375-9474(97)00596-4},
	adsurl       = {https://ui.adsabs.harvard.edu/abs/1997NuPhA.627..710C}
}

@article{Chen2009,
	title        = {Higher-order effects on the incompressibility of isospin asymmetric nuclear matter},
	author       = {Chen, Lie-Wen and Cai, Bao-Jun and Ko, Che Ming and Li, Bao-An and Shen, Chun and Xu, Jun},
	year         = 2009,
	month        = jul,
	journal      = {Phys. Rev. C},
	publisher    = {American Physical Society},
	volume       = 80,
	pages        = {014322},
	doi          = {10.1103/PhysRevC.80.014322},
	url          = {https://link.aps.org/doi/10.1103/PhysRevC.80.014322},
	issue        = 1,
	numpages     = 24
}

@article{Chen2010,
	title        = {Density slope of the nuclear symmetry energy from the neutron skin thickness of heavy nuclei},
	author       = {{Chen}, Lie-Wen and {Ko}, Che Ming and {Li}, Bao-An and {Xu}, Jun},
	year         = 2010,
	month        = aug,
	journal      = {\prc},
	volume       = 82,
	number       = 2,
	pages        = {024321},
	doi          = {10.1103/PhysRevC.82.024321},
	eid          = {024321},
	archiveprefix = {arXiv},
	eprint       = {1004.4672},
	primaryclass = {nucl-th},
	adsurl       = {https://ui.adsabs.harvard.edu/abs/2010PhRvC..82b4321C}
}

@article{Choudhury2024,
	title        = {{A NICER View of the Nearest and Brightest Millisecond Pulsar: PSR J0437{\textendash}4715}},
	author       = {{Choudhury}, Devarshi and {Salmi}, Tuomo and {Vinciguerra}, Serena and {Riley}, Thomas E. and {Kini}, Yves and {Watts}, Anna L. and {Dorsman}, Bas and {Bogdanov}, Slavko and {Guillot}, Sebastien and {Ray}, Paul S. and {Reardon}, Daniel J. and {Remillard}, Ronald A. and {Bilous}, Anna V. and {Huppenkothen}, Daniela and {Lattimer}, James M. and {Rutherford}, Nathan and {Arzoumanian}, Zaven and {Gendreau}, Keith C. and {Morsink}, Sharon M. and {Ho}, Wynn C.~G.},
	year         = 2024,
	month        = aug,
	journal      = {The Astrophysical Journal Letters},
	volume       = 971,
	number       = 1,
	pages        = {L20},
	doi          = {10.3847/2041-8213/ad5a6f},
	eid          = {L20},
	archiveprefix = {arXiv},
	eprint       = {2407.06789},
	primaryclass = {astro-ph.HE},
	adsurl       = {https://ui.adsabs.harvard.edu/abs/2024ApJ...971L..20C}
}

@article{Colo2013,
	title        = {Self-consistent RPA calculations with Skyrme-type interactions: The skyrme\_rpa program},
	author       = {{Col{\`o}}, Gianluca and {Cao}, Ligang and {Van Giai}, Nguyen and {Capelli}, Luigi},
	year         = 2013,
	month        = jan,
	journal      = {Computer Physics Communications},
	volume       = 184,
	number       = 1,
	pages        = {142--161},
	doi          = {10.1016/j.cpc.2012.07.016},
	adsurl       = {https://ui.adsabs.harvard.edu/abs/2013CoPhC.184..142C}
}

@article{Colo2021,
	title        = {User guide for the hfbcs-qrpa(v1) code},
	author       = {{Col{\`o}}, Gianluca and {Roca-Maza}, Xavier},
	year         = 2021,
	month        = feb,
	journal      = {arXiv e-prints},
	pages        = {arXiv:2102.06562},
	doi          = {10.48550/arXiv.2102.06562},
	eid          = {arXiv:2102.06562},
	archiveprefix = {arXiv},
	eprint       = {2102.06562},
	primaryclass = {nucl-th},
	adsurl       = {https://ui.adsabs.harvard.edu/abs/2021arXiv210206562C}
}

@article{Fonseca_2021,
	title        = {Refined Mass and Geometric Measurements of the High-mass PSR J0740+6620},
	author       = {Fonseca, E. and Cromartie, H. T. and Pennucci, T. T. and Ray, P. S. and Kirichenko, A. Yu. and Ransom, S. M. and Demorest, P. B. and Stairs, I. H. and Arzoumanian, Z. and Guillemot, L. and Parthasarathy, A. and Kerr, M. and Cognard, I. and Baker, P. T. and Blumer, H. and Brook, P. R. and DeCesar, M. and Dolch, T. and Dong, F. A. and Ferrara, E. C. and Fiore, W. and Garver-Daniels, N. and Good, D. C. and Jennings, R. and Jones, M. L. and Kaspi, V. M. and Lam, M. T. and Lorimer, D. R. and Luo, J. and McEwen, A. and McKee, J. W. and McLaughlin, M. A. and McMann, N. and Meyers, B. W. and Naidu, A. and Ng, C. and Nice, D. J. and Pol, N. and Radovan, H. A. and Shapiro-Albert, B. and Tan, C. M. and Tendulkar, S. P. and Swiggum, J. K. and Wahl, H. M. and Zhu, W. W.},
	year         = 2021,
	month        = jul,
	journal      = {The Astrophysical Journal Letters},
	publisher    = {The American Astronomical Society},
	volume       = 915,
	number       = 1,
	pages        = {L12},
	doi          = {10.3847/2041-8213/ac03b8},
	url          = {https://doi.org/10.3847/2041-8213/ac03b8}
}

@article{Gupta2016,
	author       = {Y.K. Gupta and U. Garg and K.B. Howard and J.T. Matta and M. Şenyiğit and M. Itoh and S. Ando and T. Aoki and A. Uchiyama and S. Adachi and M. Fujiwara and C. Iwamoto and A. Tamii and H. Akimune and C. Kadono and Y. Matsuda and T. Nakahara and T. Furuno and T. Kawabata and M. Tsumura and M.N. Harakeh and N. Kalantar-Nayestanaki},
	year         = 2016,
	volume       = 760,
	pages        = {482--485},
	doi          = {https://doi.org/10.1016/j.physletb.2016.07.021},
	issn         = {0370-2693},
	url          = {https://www.sciencedirect.com/science/article/pii/S0370269316303628}
}

@article{Gupta2018,
	title        = {Isoscalar giant monopole, dipole, and quadrupole resonances in $^{90,92}\mathbf{Zr}$ and $^{92}\mathbf{Mo}$},
	author       = {Gupta, Y. K. and Howard, K. B. and Garg, U. and Matta, J. T. and \ifmmode \mbox{\c{S}}\else \c{S}\fi{}enyi\ifmmode \breve{g}\else \u{g}\fi{}it, M. and Itoh, M. and Ando, S. and Aoki, T. and Uchiyama, A. and Adachi, S. and Fujiwara, M. and Iwamoto, C. and Tamii, A. and Akimune, H. and Kadono, C. and Matsuda, Y. and Nakahara, T. and Furuno, T. and Kawabata, T. and Tsumura, M. and Harakeh, M. N. and Kalantar-Nayestanaki, N.},
	year         = 2018,
	month        = jun,
	journal      = {Phys. Rev. C},
	publisher    = {American Physical Society},
	volume       = 97,
	pages        = {064323},
	doi          = {10.1103/PhysRevC.97.064323},
	url          = {https://link.aps.org/doi/10.1103/PhysRevC.97.064323},
	issue        = 6,
	numpages     = 11
}

@article{Klausner2025,
	title        = {Impact of ground-state properties and collective excitations on the Skyrme ansatz: A Bayesian study},
	author       = {Klausner, Pietro and Col\`o, Gianluca and Roca-Maza, Xavier and Vigezzi, Enrico},
	year         = 2025,
	month        = jan,
	journal      = {Phys. Rev. C},
	publisher    = {American Physical Society},
	volume       = 111,
	pages        = {014311},
	doi          = {10.1103/PhysRevC.111.014311},
	url          = {https://link.aps.org/doi/10.1103/PhysRevC.111.014311},
	issue        = 1,
	numpages     = 14
}

@article{Klausner2025_2,
	title        = {Properties of the neutron star crust informed by nuclear structure data},
	author       = {Klausner, Pietro and Antonelli, Marco and Gulminelli, Francesca},
	year         = 2026,
	month        = feb,
	journal      = {Phys. Rev. C},
	publisher    = {American Physical Society},
	volume       = 113,
	pages        = {025808},
	doi          = {10.1103/mm6h-3jqs},
	url          = {https://link.aps.org/doi/10.1103/mm6h-3jqs},
	issue        = 2,
	numpages     = 18
}

@misc{madai,
note="http://madai.phy.duke.edu",
url = {http://madai.phy.duke.edu}
}

@article{Margueron2018,
	title        = {Equation of state for dense nucleonic matter from metamodeling. I. Foundational aspects},
	author       = {{Margueron}, J{\'e}r{\^o}me and {Hoffmann Casali}, Rudiney and {Gulminelli}, Francesca},
	year         = 2018,
	month        = feb,
	journal      = {Phys. Rev. C},
	volume       = 97,
	number       = 2,
	pages        = {025805},
	doi          = {10.1103/PhysRevC.97.025805},
	eid          = {025805},
	archiveprefix = {arXiv},
	eprint       = {1708.06894},
	primaryclass = {nucl-th},
	adsurl       = {https://ui.adsabs.harvard.edu/abs/2018PhRvC..97b5805M}
}

@article{Mauviard2025,
	title        = {A NICER view of the 1.4 solar-mass edge-on pulsar PSR J0614--3329},
	author       = {{Mauviard}, Lucien and {Guillot}, Sebastien and {Salmi}, Tuomo and {Choudhury}, Devarshi and {Dorsman}, Bas and {Gonz{\'a}lez-Caniulef}, Denis and {Hoogkamer}, Mariska and {Huppenkothen}, Daniela and {Kazantsev}, Christine and {Kini}, Yves and {Olive}, Jean-Francois and {Stammler}, Pierre and {Watts}, Anna L. and {Mendes}, Melissa and {Rutherford}, Nathan and {Schwenk}, Achim and {Svensson}, Isak and {Bogdanov}, Slavko and {Kerr}, Matthew and {Ray}, Paul S. and {Guillemot}, Lucas and {Cognard}, Isma{\"e}l and {Theureau}, Gilles},
	year         = 2025,
	month        = jun,
	journal      = {arXiv e-prints},
	pages        = {arXiv:2506.14883},
	doi          = {10.48550/arXiv.2506.14883},
	eid          = {arXiv:2506.14883},
	archiveprefix = {arXiv},
	eprint       = {2506.14883},
	primaryclass = {astro-ph.HE},
	adsurl       = {https://ui.adsabs.harvard.edu/abs/2025arXiv250614883M}
}

@article{Miller2021,
	title        = {The Radius of PSR J0740+6620 from NICER and XMM-Newton Data},
	author       = {{Miller}, M.~C. and {Lamb}, F.~K. and {Dittmann}, A.~J. and {Bogdanov}, S. and {Arzoumanian}, Z. and {Gendreau}, K.~C. and {Guillot}, S. and {Ho}, W.~C.~G. and {Lattimer}, J.~M. and {Loewenstein}, M. and {Morsink}, S.~M. and {Ray}, P.~S. and {Wolff}, M.~T. and {Baker}, C.~L. and {Cazeau}, T. and {Manthripragada}, S. and {Markwardt}, C.~B. and {Okajima}, T. and {Pollard}, S. and {Cognard}, I. and {Cromartie}, H.~T. and {Fonseca}, E. and {Guillemot}, L. and {Kerr}, M. and {Parthasarathy}, A. and {Pennucci}, T.~T. and {Ransom}, S. and {Stairs}, I.},
	year         = 2021,
	month        = sep,
	journal      = {The Astrophysical Journal Letters},
	volume       = 918,
	number       = 2,
	pages        = {L28},
	doi          = {10.3847/2041-8213/ac089b},
	eid          = {L28},
	archiveprefix = {arXiv},
	eprint       = {2105.06979},
	primaryclass = {astro-ph.HE},
	adsurl       = {https://ui.adsabs.harvard.edu/abs/2021ApJ...918L..28M}
}

@article{monte2025aa,
	title        = {{Frozen and {\ensuremath{\beta}}-equilibrated f and p modes of cold neutron stars: Nuclear metamodel predictions}},
	author       = {{Montefusco}, Gabriele and {Antonelli}, Marco and {Gulminelli}, Francesca},
	year         = 2025,
	month        = feb,
	journal      = {\aap},
	volume       = 694,
	pages        = {A150},
	doi          = {10.1051/0004-6361/202452727},
	eid          = {A150},
	archiveprefix = {arXiv},
	eprint       = {2410.08008},
	primaryclass = {nucl-th},
	adsurl       = {https://ui.adsabs.harvard.edu/abs/2025A&A...694A.150M}
}

@article{Pearson2018,
	title        = {Unified equations of state for cold non-accreting neutron stars with Brussels–Montreal functionals – I. Role of symmetry energy},
	author       = {Pearson, J M and Chamel, N and Potekhin, A Y and Fantina, A F and Ducoin, C and Dutta, A K and Goriely, S},
	year         = 2018,
	month        = {09},
	journal      = {Monthly Notices of the Royal Astronomical Society},
	volume       = 481,
	number       = 3,
	pages        = {2994--3026},
	doi          = {10.1093/mnras/sty2413},
	issn         = {0035-8711},
	url          = {https://doi.org/10.1093/mnras/sty2413},
	eprint       = {https://academic.oup.com/mnras/article-pdf/481/3/2994/25817956/sty2413.pdf}
}

@book{Rasmussen_book,
	title        = {Gaussian Processes for Machine Learning},
	author       = {{Rasmussen}, Carl Edward and {Williams}, Christopher K.~I.},
	year         = 2006,
	publisher    = {MIT Press},
	adsurl       = {https://ui.adsabs.harvard.edu/abs/2006gpml.book.....R}
}

@article{Reinhard2022,
	title        = {Combined Theoretical Analysis of the Parity-Violating Asymmetry for $^{48}\mathrm{Ca}$ and $^{208}\mathrm{Pb}$},
	author       = {Reinhard, Paul-Gerhard and Roca-Maza, Xavier and Nazarewicz, Witold},
	year         = 2022,
	month        = dec,
	journal      = {Phys. Rev. Lett.},
	publisher    = {American Physical Society},
	volume       = 129,
	pages        = 232501,
	doi          = {10.1103/PhysRevLett.129.232501},
	url          = {https://link.aps.org/doi/10.1103/PhysRevLett.129.232501},
	issue        = 23,
	numpages     = 7
}

@article{Roca-Maza2013_2,
	title        = {Electric dipole polarizability in ${}^{208}${Pb}: {Insights} from the droplet model},
	author       = {Roca-Maza, X. and Brenna, M. and Col\`o, G. and Centelles, M. and Vi\~nas, X. and Agrawal, B. K. and Paar, N. and Vretenar, D. and Piekarewicz, J.},
	year         = 2013,
	month        = aug,
	journal      = {Phys. Rev. C},
	publisher    = {American Physical Society},
	volume       = 88,
	pages        = {024316},
	doi          = {10.1103/PhysRevC.88.024316},
	url          = {https://link.aps.org/doi/10.1103/PhysRevC.88.024316},
	issue        = 2,
	numpages     = 7
}

@article{Roca-Maza2015,
	title        = {Neutron skin thickness from the measured electric dipole polarizability in $^{68}\text{Ni}$, $^{120}\text{Sn}$, and $^{208}\text{Pb}$},
	author       = {Roca-Maza, X. and Vi\~nas, X. and Centelles, M. and Agrawal, B. K. and Col\`o, G. and Paar, N. and Piekarewicz, J. and Vretenar, D.},
	year         = 2015,
	month        = dec,
	journal      = {Phys. Rev. C},
	publisher    = {American Physical Society},
	volume       = 92,
	pages        = {064304},
	doi          = {10.1103/PhysRevC.92.064304},
	url          = {https://link.aps.org/doi/10.1103/PhysRevC.92.064304},
	issue        = 6,
	numpages     = 11
}

@article{Roca2018,
	title        = {Nuclear equation of state from ground and collective excited state properties of nuclei},
	author       = {X. Roca-Maza and N. Paar},
	year         = 2018,
	journal      = {Progress in Particle and Nuclear Physics},
	volume       = 101,
	pages        = {96--176},
	doi          = {https://doi.org/10.1016/j.ppnp.2018.04.001},
	issn         = {0146-6410},
	url          = {https://www.sciencedirect.com/science/article/pii/S0146641018300334}
}

@article{Saffer_2025,
	title        = {A Lower Mass Estimate for PSR J0348+0432 Based on CHIME/Pulsar Precision Timing},
	author       = {{Saffer}, Alexander and {Fonseca}, Emmanuel and {Ransom}, Scott and {Stairs}, Ingrid and {Lynch}, Ryan and {Good}, Deborah and {Masui}, Kiyoshi W. and {McKee}, James W. and {Meyers}, Bradley W. and {Patil}, Swarali Shivraj and {Tan}, Chia Min},
	year         = 2025,
	month        = apr,
	journal      = {\apjl},
	volume       = 983,
	number       = 1,
	pages        = {L20},
	doi          = {10.3847/2041-8213/adc25e},
	eid          = {L20},
	archiveprefix = {arXiv},
	eprint       = {2412.02850},
	primaryclass = {astro-ph.HE},
	adsurl       = {https://ui.adsabs.harvard.edu/abs/2025ApJ...983L..20S}
}

@article{Satula1997,
	title        = {Competition between T = 0 and T = 1 pairing in proton-rich nuclei},
	author       = {W. Satuła and R. Wyss},
	year         = 1997,
	journal      = {Physics Letters B},
	volume       = 393,
	number       = 1,
	pages        = {1--6},
	doi          = {https://doi.org/10.1016/S0370-2693(96)01603-6},
	issn         = {0370-2693},
	url          = {https://www.sciencedirect.com/science/article/pii/S0370269396016036}
}

@article{Satula1997_2,
	title        = {On the origin of the Wigner energy},
	author       = {W. Satuła and D.J. Dean and J. Gary and S. Mizutori and W. Nazarewicz},
	year         = 1997,
	journal      = {Physics Letters B},
	volume       = 407,
	number       = 2,
	pages        = {103--109},
	doi          = {https://doi.org/10.1016/S0370-2693(97)00711-9},
	issn         = {0370-2693},
	url          = {https://www.sciencedirect.com/science/article/pii/S0370269397007119}
}

@article{Satula1998,
	title        = {Odd-Even Staggering of Nuclear Masses: Pairing or Shape Effect?},
	author       = {Satu\l{}a, W. and Dobaczewski, J. and Nazarewicz, W.},
	year         = 1998,
	month        = oct,
	journal      = {Phys. Rev. Lett.},
	publisher    = {American Physical Society},
	volume       = 81,
	pages        = {3599--3602},
	doi          = {10.1103/PhysRevLett.81.3599},
	url          = {https://link.aps.org/doi/10.1103/PhysRevLett.81.3599},
	issue        = 17,
	numpages     = {0}
}

@article{Satula2000,
	title        = {A number projected model with generalized pairing interaction},
	author       = {W. Satuła and R. Wyss},
	year         = 2000,
	journal      = {Nuclear Physics A},
	volume       = 676,
	number       = 1,
	pages        = {120--142},
	doi          = {https://doi.org/10.1016/S0375-9474(00)00222-0},
	issn         = {0375-9474},
	url          = {https://www.sciencedirect.com/science/article/pii/S0375947400002220}
}

@book{Schunck_book,
	title        = {Energy Density Functional Methods for Atomic Nuclei},
	year         = 2019,
	publisher    = {IOP Publishing},
	series       = {2053-2563},
	doi          = {10.1088/2053-2563/aae0ed},
	isbn         = {978-0-7503-1422-0},
	url          = {https://doi.org/10.1088/2053-2563/aae0ed},
	editor       = {Schunck, Nicolas}
}

@article{Sotani2024,
	title        = {Shear oscillations in neutron stars and the nuclear symmetry energy},
	author       = {{Sotani}, Hajime},
	year         = 2024,
	month        = jan,
	journal      = {\prd},
	volume       = 109,
	number       = 2,
	pages        = {023030},
	doi          = {10.1103/PhysRevD.109.023030},
	eid          = {023030},
	archiveprefix = {arXiv},
	eprint       = {2401.08382},
	primaryclass = {astro-ph.HE},
	adsurl       = {https://ui.adsabs.harvard.edu/abs/2024PhRvD.109b3030S}
}

@article{Vinciguerra2024,
	title        = {An Updated Mass-Radius Analysis of the 2017-2018 NICER Data Set of PSR J0030+0451},
	author       = {{Vinciguerra}, Serena and {Salmi}, Tuomo and {Watts}, Anna L. and {Choudhury}, Devarshi and {Riley}, Thomas E. and {Ray}, Paul S. and {Bogdanov}, Slavko and {Kini}, Yves and {Guillot}, Sebastien and {Chakrabarty}, Deepto and {Ho}, Wynn C.~G. and {Huppenkothen}, Daniela and {Morsink}, Sharon M. and {Wadiasingh}, Zorawar and {Wolff}, Michael T.},
	year         = 2024,
	month        = jan,
	journal      = {Astrophysical Journal},
	volume       = 961,
	number       = 1,
	pages        = 62,
	doi          = {10.3847/1538-4357/acfb83},
	eid          = 62,
	archiveprefix = {arXiv},
	eprint       = {2308.09469},
	primaryclass = {astro-ph.HE},
	adsurl       = {https://ui.adsabs.harvard.edu/abs/2024ApJ...961...62V}
}

@article{Wang_2021,
	title        = {The AME 2020 atomic mass evaluation (II). Tables, graphs and references*},
	author       = {Wang, Meng and Huang, W.J. and Kondev, F.G. and Audi, G. and Naimi, S.},
	year         = 2021,
	month        = mar,
	journal      = {Chinese Physics C},
	publisher    = {Chinese Physical Society and the Institute of High Energy Physics of the Chinese Academy of Sciences and the Institute of Modern Physics of the Chinese Academy of Sciences and IOP Publishing Ltd},
	volume       = 45,
	number       = 3,
	pages        = {030003},
	doi          = {10.1088/1674-1137/abddaf},
	url          = {https://doi.org/10.1088/1674-1137/abddaf}
}

@article{Xu2022,
	title        = {Bayesian inference of finite-nuclei observables based on the KIDS model},
	author       = {{Xu}, Jun and {Papakonstantinou}, Panagiota},
	year         = 2022,
	month        = apr,
	journal      = {\prc},
	volume       = 105,
	number       = 4,
	pages        = {044305},
	doi          = {10.1103/PhysRevC.105.044305},
	eid          = {044305},
	archiveprefix = {arXiv},
	eprint       = {2201.03835},
	primaryclass = {nucl-th},
	adsurl       = {https://ui.adsabs.harvard.edu/abs/2022PhRvC.105d4305X}
}

@misc{Yue2024,
	title        = {PREX and CREX: Evidence for Strong Isovector Spin-Orbit Interaction},
	author       = {Tong-Gang Yue and Zhen Zhang and Lie-Wen Chen},
	year         = 2024,
	url          = {https://arxiv.org/abs/2406.03844},
	eprint       = {2406.03844},
	archiveprefix = {arXiv},
	primaryclass = {nucl-th}
}

@article{Zemlyakov2023,
	title        = {Neutron star inner crust: reduction of shear modulus by nuclei finite size effect},
	author       = {{Zemlyakov}, Nikita A. and {Chugunov}, Andrey I.},
	year         = 2023,
	month        = jan,
	journal      = {\mnras},
	volume       = 518,
	number       = 3,
	pages        = {3813--3819},
	doi          = {10.1093/mnras/stac3377},
	archiveprefix = {arXiv},
	eprint       = {2209.05821},
	primaryclass = {astro-ph.HE},
	adsurl       = {https://ui.adsabs.harvard.edu/abs/2023MNRAS.518.3813Z}
}

@article{Zemlyakov2025,
	title        = {Constraining the shear modulus of a polycrystalline neutron star crust: Hashin-Shtrikman variational approach},
	author       = {Zemlyakov, Nikita A. and Chugunov, Andrey I.},
	year         = 2025,
	month        = aug,
	journal      = {Phys. Rev. D},
	publisher    = {American Physical Society},
	volume       = 112,
	pages        = {043032},
	doi          = {10.1103/lqd6-56zn},
	url          = {https://link.aps.org/doi/10.1103/lqd6-56zn},
	issue        = 4,
	numpages     = 12
}

@article{RingSchuck,
	title        = {The Nuclear Many-Body Problem},
	author       = {{Ring}, P. and {Schuck}, P. and {Strayer}, M.~R.},
	year         = 1983,
	month        = jan,
	journal      = {Physics Today},
	volume       = 36,
	number       = 7,
	pages        = 70,
	doi          = {10.1063/1.2915762},
	adsurl       = {https://ui.adsabs.harvard.edu/abs/1983PhT....36g..70R}
}

@misc{Papakonstantinou2026,
	title        = {CREX and PREX-II reconciled within energy-density functional theory},
	author       = {P. Papakonstantinou},
	year         = 2026,
	url          = {https://arxiv.org/abs/2602.22722},
	eprint       = {2602.22722},
	archiveprefix = {arXiv},
	primaryclass = {nucl-th}
}

@article{Qiu2025,
title = {A relativistic mechanism for the enhanced isovector spin-orbit interaction suggested by parity-violating electron scattering experiments},
journal = {Physics Letters B},
volume = {879},
pages = {140648},
year = {2026},
issn = {0370-2693},
doi = {https://doi.org/10.1016/j.physletb.2026.140648},
url = {https://www.sciencedirect.com/science/article/pii/S0370269326005009},
author = {Mengying Qiu and Tong-Gang Yue and Zhen Zhang and Lie-Wen Chen},
}

@article{Zhao2025,
	title        = {Characterizing the nuclear models informed by PREX and CREX: A view from Bayesian inference},
	author       = {Zhao, Tianqi and Lin, Zidu and Kumar, Bharat and Steiner, Andrew W. and Prakash, Madappa},
	year         = 2025,
	month        = dec,
	journal      = {Phys. Rev. Res.},
	publisher    = {American Physical Society},
	volume       = 7,
	pages        = {043335},
	doi          = {10.1103/472x-9cxj},
	url          = {https://link.aps.org/doi/10.1103/472x-9cxj},
	issue        = 4,
	numpages     = 20
}

@article{Salinas2024,
	title        = {Impact of tensor couplings with scalar mixing on covariant energy density functionals},
	author       = {Salinas, Marc and Piekarewicz, J.},
	year         = 2024,
	month        = apr,
	journal      = {Phys. Rev. C},
	publisher    = {American Physical Society},
	volume       = 109,
	pages        = {045807},
	doi          = {10.1103/PhysRevC.109.045807},
	url          = {https://link.aps.org/doi/10.1103/PhysRevC.109.045807},
	issue        = 4,
	numpages     = 15
}

@article{RocaMaza2008,
	title        = {Impact of the symmetry energy on the outer crust of nonaccreting neutron stars},
	author       = {Roca-Maza, X. and Piekarewicz, J.},
	year         = 2008,
	month        = aug,
	journal      = {Phys. Rev. C},
	publisher    = {American Physical Society},
	volume       = 78,
	pages        = {025807},
	doi          = {10.1103/PhysRevC.78.025807},
	url          = {https://link.aps.org/doi/10.1103/PhysRevC.78.025807},
	issue        = 2,
	numpages     = 11
}

@ARTICLE{Yuskel2019,
       author = {{Y{\"u}ksel}, E. and {Marketin}, T. and {Paar}, N.},
        title = "{Optimizing the relativistic energy density functional with nuclear ground state and collective excitation properties}",
      journal = {\prc},
         year = 2019,
        month = mar,
       volume = {99},
       number = {3},
          eid = {034318},
        pages = {034318},
          doi = {10.1103/PhysRevC.99.034318},
archivePrefix = {arXiv},
       eprint = {1901.05552},
 primaryClass = {nucl-th},
       adsurl = {https://ui.adsabs.harvard.edu/abs/2019PhRvC..99c4318Y}
}

@ARTICLE{Zhang2023,
       author = {{Zhang}, Zhen and {Chen}, Lie-Wen},
        title = "{Bayesian inference of the symmetry energy and the neutron skin in $^{48}$Ca and $^{208}$Pb from CREX and PREX-2}",
      journal = {\prc},
         year = 2023,
        month = aug,
       volume = {108},
       number = {2},
          eid = {024317},
        pages = {024317},
          doi = {10.1103/PhysRevC.108.024317},
archivePrefix = {arXiv},
       eprint = {2207.03328},
 primaryClass = {nucl-th},
       adsurl = {https://ui.adsabs.harvard.edu/abs/2023PhRvC.108b4317Z}
}

@article{Scurto2024,
  title = {General predictions of neutron star properties using unified relativistic mean-field equations of state},
  author = {Scurto, Luigi and Pais, Helena and Gulminelli, Francesca},
  journal = {Phys. Rev. D},
  volume = {109},
  issue = {10},
  pages = {103015},
  numpages = {26},
  year = {2024},
  month = {May},
  publisher = {American Physical Society},
  doi = {10.1103/PhysRevD.109.103015},
  url = {https://link.aps.org/doi/10.1103/PhysRevD.109.103015}
}

@article{Char2025,
  title = {Exploring the limits of nucleonic metamodeling using different relativistic density functionals},
  author = {Char, Prasanta and Mondal, Chiranjib},
  journal = {Phys. Rev. D},
  volume = {111},
  issue = {10},
  pages = {103024},
  numpages = {12},
  year = {2025},
  month = {May},
  publisher = {American Physical Society},
  doi = {10.1103/PhysRevD.111.103024},
  url = {https://link.aps.org/doi/10.1103/PhysRevD.111.103024}
}

@Article{DinhThi2021,
AUTHOR = {Dinh Thi, Hoa and Mondal, Chiranjib and Gulminelli, Francesca},
TITLE = {The Nuclear Matter Density Functional under the Nucleonic Hypothesis},
JOURNAL = {Universe},
VOLUME = {7},
YEAR = {2021},
NUMBER = {10},
ARTICLE-NUMBER = {373},
URL = {https://www.mdpi.com/2218-1997/7/10/373},
ISSN = {2218-1997},
DOI = {10.3390/universe7100373}
}

@article{Erler2012,
	Author = {Erler, Jochen and Birge, Noah and Kortelainen, Markus and Nazarewicz, Witold and Olsen, Erik and Perhac, Alexander M. and Stoitsov, Mario},
	Journal = {Nature},
	Number = {7404},
	Pages = {509--512},
	Title = {The limits of the nuclear landscape},
	Volume = {486},
	Year = {2012}}

@article{Landry2019,
  title = {Nonparametric inference of the neutron star equation of state from gravitational wave observations},
  author = {Landry, Philippe and Essick, Reed},
  journal = {Phys. Rev. D},
  volume = {99},
  issue = {8},
  pages = {084049},
  numpages = {27},
  year = {2019},
  month = {Apr},
  publisher = {American Physical Society},
  doi = {10.1103/PhysRevD.99.084049},
  url = {https://link.aps.org/doi/10.1103/PhysRevD.99.084049}
}

@article{Beznogov2024,
  title = {Bayesian inference of the dense matter equation of state built upon extended Skyrme interactions},
  author = {Beznogov, Mikhail V. and Raduta, Adriana R.},
  journal = {Phys. Rev. C},
  volume = {110},
  issue = {3},
  pages = {035805},
  numpages = {23},
  year = {2024},
  month = {Sep},
  publisher = {American Physical Society},
  doi = {10.1103/PhysRevC.110.035805},
  url = {https://link.aps.org/doi/10.1103/PhysRevC.110.035805}
}

@article{Kumar2024,
	Author = {Kumar, Rajesh and Dexheimer, Veronica and Jahan, Johannes and Noronha, Jorge and Noronha-Hostler, Jacquelyn and Ratti, Claudia and Yunes, Nico and Nava Acuna, Angel Rodrigo and Alford, Mark and Anik, Mahmudul Hasan and Chatterjee, Debarati and Chatziioannou, Katerina and Chen, Hsin-Yu and Clevinger, Alexander and Conde, Carlos and Cruz-Camacho, Nikolas and Dore, Travis and Drischler, Christian and Elfner, Hannah and Essick, Reed and Friedenberg, David and Ghosh, Suprovo and Grefa, Joaquin and Haas, Roland and Haber, Alexander and Hammelmann, Jan and Harris, Steven and Haster, Carl-Johan and Hatsuda, Tetsuo and Hippert, Mauricio and Hirayama, Renan and Holt, Jeremy W. and Kahangirwe, Micheal and Karthein, Jamie and Kojo, Toru and Landry, Philippe and Lin, Zidu and Luzum, Matthew and Manning, Timothy Andrew and Salinas San Martin, Jordi and Miller, Cole and Most, Elias Roland and Mroczek, Debora and Muronga, Azwinndini and Patino, Nicolas and Peterson, Jeffrey and Plumberg, Christopher and Price, Damien and Providencia, Constanca and Rougemont, Romulo and Roy, Satyajit and Shah, Hitansh and Shapiro, Stuart and Steiner, Andrew W. and Strickland, Michael and Tan, Hung and Togashi, Hajime and Portillo Vazquez, Israel and Wen, Pengsheng and Zhang, Ziyuan and MUSES Collaboration},
	Journal = {Living Reviews in Relativity},
	Number = {1},
	Pages = {3},
	Title = {Theoretical and experimental constraints for the equation of state of dense and hot matter},
	Volume = {27},
	Year = {2024},
	doi = {https://doi.org/10.1007/s41114-024-00049-6},
	url = {https://link.springer.com/article/10.1007/s41114-024-00049-6}
}

@article{Essick2020,
  title = {Nonparametric inference of neutron star composition, equation of state, and maximum mass with GW170817},
  author = {Essick, Reed and Landry, Philippe and Holz, Daniel E.},
  journal = {Phys. Rev. D},
  volume = {101},
  issue = {6},
  pages = {063007},
  numpages = {23},
  year = {2020},
  month = {Mar},
  publisher = {American Physical Society},
  doi = {10.1103/PhysRevD.101.063007},
  url = {https://link.aps.org/doi/10.1103/PhysRevD.101.063007}
}

@article{Read2009,
  title = {Constraints on a phenomenologically parametrized neutron-star equation of state},
  author = {Read, Jocelyn S. and Lackey, Benjamin D. and Owen, Benjamin J. and Friedman, John L.},
  journal = {Phys. Rev. D},
  volume = {79},
  issue = {12},
  pages = {124032},
  numpages = {18},
  year = {2009},
  month = {Jun},
  publisher = {American Physical Society},
  doi = {10.1103/PhysRevD.79.124032},
  url = {https://link.aps.org/doi/10.1103/PhysRevD.79.124032}
}

@article{Suleiman2022,
  title = {Polytropic fits of modern and unified equations of state},
  author = {Suleiman, Lami and Fortin, Morgane and Zdunik, Julian Leszek and Provid\^encia, Constan\ifmmode \mbox{\c{c}}\else \c{c}\fi{}a},
  journal = {Phys. Rev. C},
  volume = {106},
  issue = {3},
  pages = {035805},
  numpages = {22},
  year = {2022},
  month = {Sep},
  publisher = {American Physical Society},
  doi = {10.1103/PhysRevC.106.035805},
  url = {https://link.aps.org/doi/10.1103/PhysRevC.106.035805}
}

@article{Grams2026,
title = {Skyrme-Hartree-Fock-Bogoliubov mass models on a 3D mesh: V. The N2LO extension of the Skyrme EDF},
journal = {Physics Letters B},
volume = {879},
pages = {140590},
year = {2026},
issn = {0370-2693},
doi = {https://doi.org/10.1016/j.physletb.2026.140590},
url = {https://www.sciencedirect.com/science/article/pii/S0370269326004429},
author = {G. Grams and W. Ryssens and A. Sánchez-Fernández and N.N. Shchechilin and L. {González-Miret Zaragoza} and P. Demol and N. Chamel and S. Goriely and M. Bender},
}

@article{Grams2025,
	Author = {Grams, Guilherme and Shchechilin, Nikolai N. and S{\'a}nchez-Fern{\'a}ndez, Adri{\'a}n and Ryssens, Wouter and Chamel, Nicolas and Goriely, Stephane},
	Journal = {The European Physical Journal A},
	Number = {2},
	Pages = {35},
	Title = {Skyrme--Hartree--Fock--Bogoliubov mass models on a 3D mesh: IV. Improved description of the isospin dependence of pairing},
	Volume = {61},
	Year = {2025},
	doi = {https://doi.org/10.1140/epja/s10050-025-01503-x},
	url = {https://link.springer.com/article/10.1140/epja/s10050-025-01503-x#citeas}
}

@ARTICLE{Kortelainen2010,
       author = {{Kortelainen}, M. and {Lesinski}, T. and {Mor{\'e}}, J. and {Nazarewicz}, W. and {Sarich}, J. and {Schunck}, N. and {Stoitsov}, M.~V. and {Wild}, S.},
        title = "{Nuclear energy density optimization}",
      journal = {\prc},
         year = 2010,
        month = aug,
       volume = {82},
       number = {2},
          eid = {024313},
        pages = {024313},
          doi = {10.1103/PhysRevC.82.024313},
archivePrefix = {arXiv},
       eprint = {1005.5145},
 primaryClass = {nucl-th},
       adsurl = {https://ui.adsabs.harvard.edu/abs/2010PhRvC..82b4313K}
}

@ARTICLE{Moller2012,
       author = {{M{\"o}ller}, Peter and {Myers}, William D. and {Sagawa}, Hiroyuki and {Yoshida}, Satoshi},
        title = "{New Finite-Range Droplet Mass Model and Equation-of-State Parameters}",
      journal = {\prl},
         year = 2012,
        month = feb,
       volume = {108},
       number = {5},
          eid = {052501},
        pages = {052501},
          doi = {10.1103/PhysRevLett.108.052501},
       adsurl = {https://ui.adsabs.harvard.edu/abs/2012PhRvL.108e2501M}
}

@ARTICLE{Shlomo2006,
       author = {{Shlomo}, S. and {Kolomietz}, V.~M. and {Col{\`o}}, G.},
        title = "{Deducing the nuclear-matter incompressibility coefficient from data on isoscalar compression modes}",
      journal = {Eur. Phys. Jou. A},
         year = 2006,
        month = oct,
       volume = {30},
       number = {1},
        pages = {23-30},
          doi = {10.1140/epja/i2006-10100-3},
       adsurl = {https://ui.adsabs.harvard.edu/abs/2006EPJA...30...23S}
}

@ARTICLE{Khan2012,
       author = {{Khan}, E. and {Margueron}, J. and {Vida{\~n}a}, I.},
        title = "{Constraining the Nuclear Equation of State at Subsaturation Densities}",
      journal = {\prl},
         year = 2012,
        month = aug,
       volume = {109},
       number = {9},
          eid = {092501},
        pages = {092501},
          doi = {10.1103/PhysRevLett.109.092501},
archivePrefix = {arXiv},
       eprint = {1204.0399},
 primaryClass = {nucl-th},
       adsurl = {https://ui.adsabs.harvard.edu/abs/2012PhRvL.109i2501K}
}

@article{Fayans1998,
title = {Towards a universal nuclear density functional},
journal = {Journal of Experimental and Theoretical Physics Letters},
volume = {68},
number = {3},
pages = {169-174},
year = {1998},
doi = {10.1134/1.567841},
url = {https://doi.org/10.1134/1.567841},
author = {Fayans, S. A.}
}

@article{Gil2017,
	Author = {{Gil}, H. and {Oh}, Y. and {Hyun}, C.~H. and  {Papakonstantinou}, P.}, 
	Date = {2017/04/},
	Doi = {10.3938/NPSM.67.456},
	Et = {2017/04/28},
	Isbn = {0374-4914; 2289-0041},
	J2 = {New Phys.: Sae Mulli},
	Journal = {New Physics: Sae Mulli},
	La = {eng},
	Month = {04},
	Number = {4},
	Pages = {456--461},
	Publisher = {New Physics: Sae Mulli},
	Title = {Skyrme-Type Nuclear Force for the KIDS Energy Density Functional},
	Ty = {JOUR},
	Url = {http://www.npsm-kps.org/journal/view.html?doi=10.3938/NPSM.67.456},
	Volume = {67},
	Year = {2017}}

@ARTICLE{Papakonstantinou2018,
       author = {{Papakonstantinou}, Panagiota and {Park}, Tae-Sun and {Lim}, Yeunhwan and {Hyun}, Chang Ho},
        title = "{Density dependence of the nuclear energy-density functional}",
      journal = {\prc},
         year = 2018,
        month = jan,
       volume = {97},
       number = {1},
          eid = {014312},
        pages = {014312},
          doi = {10.1103/PhysRevC.97.014312},
archivePrefix = {arXiv},
       eprint = {1606.04219},
 primaryClass = {nucl-th},
       adsurl = {https://ui.adsabs.harvard.edu/abs/2018PhRvC..97a4312P}
}

@ARTICLE{Gil2017_ActaPolonica,
       author = {{Gil}, H. and {Papakonstantinou}, P. and {Hyun}, C.~H. and {Park}, T. -S. and {Oh}, Y.},
        title = "{Nuclear Energy Density Functional for KIDS}",
      journal = {Acta Physica Polonica B},
         year = 2017,
        month = jan,
       volume = {48},
       number = {3},
        pages = {305},
          doi = {10.5506/APhysPolB.48.305},
archivePrefix = {arXiv},
       eprint = {1611.04257},
 primaryClass = {nucl-th},
       adsurl = {https://ui.adsabs.harvard.edu/abs/2017AcPPB..48..305G}
}

@article{Rutherford2024,
doi = {10.3847/2041-8213/ad5f02},
url = {https://doi.org/10.3847/2041-8213/ad5f02},
year = {2024},
month = {aug},
publisher = {The American Astronomical Society},
volume = {971},
number = {1},
pages = {L19},
author = {Rutherford, Nathan and Mendes, Melissa and Svensson, Isak and Schwenk, Achim and Watts, Anna L. and Hebeler, Kai and Keller, Jonas and Prescod-Weinstein, Chanda and Choudhury, Devarshi and Raaijmakers, Geert and Salmi, Tuomo and Timmerman, Patrick and Vinciguerra, Serena and Guillot, Sebastien and Lattimer, James M.},
title = {Constraining the Dense Matter Equation of State with New NICER Mass–Radius Measurements and New Chiral Effective Field Theory Inputs},
journal = {The Astrophysical Journal Letters},
}

@article{Melendez2019,
  title = {Quantifying correlated truncation errors in effective field theory},
  author = {Melendez, J. A. and Furnstahl, R. J. and Phillips, D. R. and Pratola, M. T. and Wesolowski, S.},
  journal = {Phys. Rev. C},
  volume = {100},
  issue = {4},
  pages = {044001},
  numpages = {22},
  year = {2019},
  month = {Oct},
  publisher = {American Physical Society},
  doi = {10.1103/PhysRevC.100.044001},
  url = {https://link.aps.org/doi/10.1103/PhysRevC.100.044001}
}

@misc{suppmaterial,
  title = {See Supplementary material at \href{https://arxiv.org/abs/2604.11358}{https://arxiv.org/abs/2604.11358} for files containing numerical values of Tables V and VI},
}

\appendix

\section{Mean vector and covariance matrix}
\label{app:gausian_fit_tables}

We approximate the posterior sample of EDF parameters (or derived nuclear-matter parameters) with a single multivariate normal distribution $\mathcal{N}(\mathbf{x}|\boldsymbol{\mu},\boldsymbol{\Sigma})$ by matching the posterior's sample mean and covariance, see Eq.~\eqref{eq:multivariate_Gaussian}. 
This provides a compact representation of the full posterior that allows downstream users to propagate uncertainties using only $\boldsymbol{\mu}$ and $\boldsymbol{\Sigma}$. 
This compression of the full posterior information into $\boldsymbol{\mu}$ and $\boldsymbol{\Sigma}$ reproduces linear correlations extremely well, but it does not capture possible skewness or multimodality features, most notably for~$G_1$ (the quality of this compression is illustrated in Figures~\ref{fig:prior_vs_gaussian_nmp} and~\ref{fig:prior_vs_gaussian_surf}).

The mean vector $\boldsymbol{\mu}$ and of the (symmetric and positive-definite) covariance matrix $\boldsymbol{\Sigma}$ are given in Tables~\ref{tab:multivariate_gaussian_mean} and~\ref{tab:multivariate_gaussian_cov}.

\begin{table}
\centering
\caption{
Mean values $\boldsymbol{\mu}$ for the 15 variables $\mathbf{x}$ in our posterior.
The marginal distribution of each $x_i$ ($i=1,...,15$) is an univariate Gaussian with standard deviation $\sigma_i = \sqrt{\Sigma_{ii}}$, which is also reported.
}

\label{tab:multivariate_gaussian_mean}
\begin{tabular}{ccccc}
\hline
\qquad\qquad & $\mathbf{x}$\qquad & $\mu_i$\qquad & $\sigma_i$\qquad & \qquad\qquad\qquad \\ 
\hline
 1 &$n_{sat}$ & $1.6 \times 10^{-1}$ & $3.2 \times 10^{-3}$ & fm$^{-3}$\\ 
 2 &$E_{sat}$ & $-1.6 \times 10^{1}$ & $9.1 \times 10^{-2}$ & MeV\\ 
 3 &$K_{sat}$ & $2.4 \times 10^{2}$ & $1.0 \times 10^{1}$ & MeV\\ 
 4 &$Q_{sat}$ & $-3.8 \times 10^{2}$ & $2.0 \times 10^{1}$ & MeV\\ 
 5 &$Z_{sat}$ & $1.6 \times 10^{3}$ & $1.8 \times 10^{2}$ & MeV\\ 
 6 &$E_{sym}$ & $3.1 \times 10^{1}$ & $1.9$ & MeV\\ 
 7 &$L_{sym}$ & $2.6 \times 10^{1}$ & $1.6 \times 10^{1}$ & MeV\\ 
 8 &$K_{sym}$ & $-2.6 \times 10^{2}$ & $5.1 \times 10^{1}$ & MeV\\ 
 9 &$Q_{sym}$ & $4.8 \times 10^{2}$ & $8.9 \times 10^{1}$ & MeV\\ 
 10 &$Z_{sym}$ & $-2.5 \times 10^{3}$ & $4.8 \times 10^{2}$ & MeV\\ 
 11 &$m^*_0/m$ & $9.1 \times 10^{-1}$ & $8.3 \times 10^{-2}$ & -\\ 
 12 &$m^*_1/m$ & $7.1 \times 10^{-1}$ & $2.1 \times 10^{-2}$ & -\\ 
 13 &$G_{0}$ & $1.3 \times 10^{2}$ & $1.1 \times 10^{1}$ & MeV fm$^{5}$\\ 
 14 &$G_{1}$ & $2.8$ & $4.4 \times 10^{1}$ & MeV fm$^{5}$\\ 
 15 &$W_{0}$ & $1.3 \times 10^{2}$ & $1.7 \times 10^{1}$ & MeV fm$^{5}$\\ 
\hline
\end{tabular}
\end{table}

\begin{table}
\centering
\caption{Covariance matrix $\boldsymbol{\Sigma}$ for the 15 variables $\mathbf{x}$ listed in Tab.~\ref{tab:multivariate_gaussian_mean}.}
\label{tab:multivariate_gaussian_cov}
\begin{tabular}{crrrrr}
\hline
$\Sigma_{ij}$ & \multicolumn{1}{c}{1} & \multicolumn{1}{c}{2}& \multicolumn{1}{c}{3}& \multicolumn{1}{c}{4}& \multicolumn{1}{c}{5} \\ 
\hline
 1 & $1.0 \times 10^{-5}$ & $-1.8 \times 10^{-4}$ & $-1.8 \times 10^{-2}$ & $-4.2 \times 10^{-2}$ & $3.8 \times 10^{-1}$ \\ 
 2 & $-1.8 \times 10^{-4}$ & $8.2 \times 10^{-3}$ & $-4.1 \times 10^{-2}$ & $2.9 \times 10^{-1}$ & $-2.1$ \\ 
 3 & $-1.8 \times 10^{-2}$ & $-4.1 \times 10^{-2}$ & $1.0 \times 10^{2}$ & $1.9 \times 10^{2}$ & $-1.8 \times 10^{3}$ \\ 
 4 & $-4.2 \times 10^{-2}$ & $2.9 \times 10^{-1}$ & $1.9 \times 10^{2}$ & $4.1 \times 10^{2}$ & $-3.6 \times 10^{3}$ \\ 
 5 & $3.8 \times 10^{-1}$ & $-2.1$ & $-1.8 \times 10^{3}$ & $-3.6 \times 10^{3}$ & $3.3 \times 10^{4}$ \\ 
 6 & $6.1 \times 10^{-4}$ & $-9.5 \times 10^{-2}$ & $2.9$ & $3.1$ & $-2.4 \times 10^{1}$ \\ 
 7 & $8.5 \times 10^{-3}$ & $-8.2 \times 10^{-1}$ & $2.9 \times 10^{1}$ & $4.3 \times 10^{1}$ & $-2.9 \times 10^{2}$ \\ 
 8 & $3.7 \times 10^{-2}$ & $-2.3$ & $2.2 \times 10^{1}$ & $9.6 \times 10^{1}$ & $1.7 \times 10^{1}$ \\ 
 9 & $-3.8 \times 10^{-2}$ & $4.0$ & $-2.6 \times 10^{2}$ & $-2.6 \times 10^{2}$ & $3.1 \times 10^{3}$ \\ 
 10 & $9.6 \times 10^{-2}$ & $-1.9 \times 10^{1}$ & $1.8 \times 10^{3}$ & $1.8 \times 10^{3}$ & $-2.3 \times 10^{4}$ \\ 
 11 & $-3.0 \times 10^{-6}$ & $8.9 \times 10^{-4}$ & $1.1 \times 10^{-2}$ & $-3.9 \times 10^{-1}$ & $5.6 \times 10^{-1}$ \\ 
 12 & $4.6 \times 10^{-7}$ & $-9.8 \times 10^{-6}$ & $-6.5 \times 10^{-4}$ & $-2.3 \times 10^{-3}$ & $1.6 \times 10^{-2}$ \\ 
 13 & $9.6 \times 10^{-4}$ & $-2.3 \times 10^{-1}$ & $-1.5 \times 10^{1}$ & $1.8 \times 10^{1}$ & $2.1 \times 10^{2}$ \\ 
 14 & $-2.3 \times 10^{-2}$ & $6.1 \times 10^{-1}$ & $2.3 \times 10^{1}$ & $6.7$ & $-4.7 \times 10^{2}$ \\ 
 15 & $-4.7 \times 10^{-3}$ & $1.0 \times 10^{-1}$ & $-9.0$ & $3.2 \times 10^{1}$ & $1.7 \times 10^{1}$ \\ 
\\ 
\end{tabular}
\begin{tabular}{crrrrr}
$\Sigma_{ij}$ & \multicolumn{1}{c}{6} & \multicolumn{1}{c}{7}& \multicolumn{1}{c}{8}& \multicolumn{1}{c}{9}& \multicolumn{1}{c}{10}  \\ 
\hline
 1  & $6.1 \times 10^{-4}$ & $8.5 \times 10^{-3}$ & $3.7 \times 10^{-2}$ & $-3.8 \times 10^{-2}$ & $9.6 \times 10^{-2}$  \\ 
 2  & $-9.5 \times 10^{-2}$ & $-8.2 \times 10^{-1}$ & $-2.3$ & $4.0$ & $-1.9 \times 10^{1}$  \\ 
 3  & $2.9$ & $2.9 \times 10^{1}$ & $2.2 \times 10^{1}$ & $-2.6 \times 10^{2}$ & $1.8 \times 10^{3}$  \\ 
 4  & $3.1$ & $4.3 \times 10^{1}$ & $9.6 \times 10^{1}$ & $-2.6 \times 10^{2}$ & $1.8 \times 10^{3}$  \\ 
 5  & $-2.4 \times 10^{1}$ & $-2.9 \times 10^{2}$ & $1.7 \times 10^{1}$ & $3.1 \times 10^{3}$ & $-2.3 \times 10^{4}$  \\ 
 6  & $3.6$ & $2.7 \times 10^{1}$ & $6.7 \times 10^{1}$ & $-1.2 \times 10^{2}$ & $5.8 \times 10^{2}$  \\ 
 7  & $2.7 \times 10^{1}$ & $2.6 \times 10^{2}$ & $7.5 \times 10^{2}$ & $-1.3 \times 10^{3}$ & $6.2 \times 10^{3}$  \\ 
 8  & $6.7 \times 10^{1}$ & $7.5 \times 10^{2}$ & $2.6 \times 10^{3}$ & $-3.4 \times 10^{3}$ & $1.5 \times 10^{4}$  \\ 
 9  & $-1.2 \times 10^{2}$ & $-1.3 \times 10^{3}$ & $-3.4 \times 10^{3}$ & $8.0 \times 10^{3}$ & $-4.2 \times 10^{4}$  \\ 
 10  & $5.8 \times 10^{2}$ & $6.2 \times 10^{3}$ & $1.5 \times 10^{4}$ & $-4.2 \times 10^{4}$ & $2.3 \times 10^{5}$  \\ 
 11  & $-3.3 \times 10^{-2}$ & $-4.1 \times 10^{-1}$ & $-2.3$ & $-4.3 \times 10^{-1}$ & $8.5$  \\ 
 12  & $-1.1 \times 10^{-4}$ & $-8.8 \times 10^{-4}$ & $-3.8 \times 10^{-3}$ & $-1.3 \times 10^{-3}$ & $1.4 \times 10^{-2}$  \\ 
 13  & $6.2$ & $6.8 \times 10^{1}$ & $3.4 \times 10^{2}$ & $-3.2 \times 10^{1}$ & $-6.8 \times 10^{2}$  \\ 
 14  & $2.6 \times 10^{1}$ & $-1.8 \times 10^{1}$ & $-5.7 \times 10^{2}$ & $3.0 \times 10^{2}$ & $-3.5 \times 10^{2}$  \\ 
 15  & $-7.4 \times 10^{-1}$ & $3.8 \times 10^{1}$ & $2.7 \times 10^{2}$ & $-4.1 \times 10^{1}$ & $-4.1 \times 10^{2}$  \\ 
\\ 
\end{tabular}
\begin{tabular}{crrrrr}
$\Sigma_{ij}$ & \multicolumn{1}{c}{11} & \multicolumn{1}{c}{12}& \multicolumn{1}{c}{13}& \multicolumn{1}{c}{14}& \multicolumn{1}{c}{15}  \\ 
\hline
 1  & $-3.0 \times 10^{-6}$ & $4.6 \times 10^{-7}$ & $9.6 \times 10^{-4}$ & $-2.3 \times 10^{-2}$ & $-4.7 \times 10^{-3}$   \\ 
 2  & $8.9 \times 10^{-4}$ & $-9.8 \times 10^{-6}$ & $-2.3 \times 10^{-1}$ & $6.1 \times 10^{-1}$ & $1.0 \times 10^{-1}$   \\ 
 3  & $1.1 \times 10^{-2}$ & $-6.5 \times 10^{-4}$ & $-1.5 \times 10^{1}$ & $2.3 \times 10^{1}$ & $-9.0$   \\ 
 4  & $-3.9 \times 10^{-1}$ & $-2.3 \times 10^{-3}$ & $1.8 \times 10^{1}$ & $6.7$ & $3.2 \times 10^{1}$   \\ 
 5  & $5.6 \times 10^{-1}$ & $1.6 \times 10^{-2}$ & $2.1 \times 10^{2}$ & $-4.7 \times 10^{2}$ & $1.7 \times 10^{1}$   \\ 
 6  & $-3.3 \times 10^{-2}$ & $-1.1 \times 10^{-4}$ & $6.2$ & $2.6 \times 10^{1}$ & $-7.4 \times 10^{-1}$   \\ 
 7  & $-4.1 \times 10^{-1}$ & $-8.8 \times 10^{-4}$ & $6.8 \times 10^{1}$ & $-1.8 \times 10^{1}$ & $3.8 \times 10^{1}$   \\ 
 8  & $-2.3$ & $-3.8 \times 10^{-3}$ & $3.4 \times 10^{2}$ & $-5.7 \times 10^{2}$ & $2.7 \times 10^{2}$   \\ 
 9  & $-4.3 \times 10^{-1}$ & $-1.3 \times 10^{-3}$ & $-3.2 \times 10^{1}$ & $3.0 \times 10^{2}$ & $-4.1 \times 10^{1}$   \\ 
 10  & $8.5$ & $1.4 \times 10^{-2}$ & $-6.8 \times 10^{2}$ & $-3.5 \times 10^{2}$ & $-4.1 \times 10^{2}$   \\ 
 11  & $6.8 \times 10^{-3}$ & $1.2 \times 10^{-5}$ & $-8.2 \times 10^{-1}$ & $9.2 \times 10^{-1}$ & $-6.1 \times 10^{-1}$   \\ 
 12  & $1.2 \times 10^{-5}$ & $4.5 \times 10^{-4}$ & $-1.2 \times 10^{-3}$ & $-4.2 \times 10^{-3}$ & $7.4 \times 10^{-5}$   \\ 
 13  & $-8.2 \times 10^{-1}$ & $-1.2 \times 10^{-3}$ & $1.2 \times 10^{2}$ & $-1.9 \times 10^{2}$ & $1.1 \times 10^{2}$   \\ 
 14  & $9.2 \times 10^{-1}$ & $-4.2 \times 10^{-3}$ & $-1.9 \times 10^{2}$ & $1.9 \times 10^{3}$ & $-5.4 \times 10^{2}$   \\ 
 15  & $-6.1 \times 10^{-1}$ & $7.4 \times 10^{-5}$ & $1.1 \times 10^{2}$ & $-5.4 \times 10^{2}$ & $2.8 \times 10^{2}$   \\ 
\hline
\end{tabular}
\end{table}

\end{document}